%% file: manuscript.tex
\documentclass[12pt,a4paper]{article}
\usepackage[utf8]{inputenc}
\usepackage[margin=1in]{geometry}
\usepackage{amsmath} 
\usepackage{amsfonts}
\usepackage{amssymb}
\usepackage{fullpage}
\usepackage{setspace}
\usepackage{lscape} 
\usepackage{graphicx}
\usepackage{float}
\usepackage{breakcites}
\usepackage{pdfpages}
\usepackage{xcolor}
\usepackage{natbib}
\usepackage{hyperref}
\usepackage{setspace}
\usepackage{multirow}
\usepackage{lineno}
\input{preamble.tex}

\title{A Class of Models for Large Zero-inflated Spatial Data}
\author{Ben Seiyon Lee\thanks{Corresponding Author: Department of Statistics, George Mason University; 4400 University Drive, MS 4A7; Fairfax, VA 22030-4444; United States of America (e-mail: \href{mailto:slee287@gmu.edu}{slee287@gmu.edu})}\\ Department of Statistics, George Mason University,\\
Fairfax, VA, United States
\and  Murali Haran\\
Department of Statistics, The Pennsylvania State University,\\
University Park, PA, United States}

\date{}
\graphicspath{{Figures/}}

\begin{document}
\pagenumbering{gobble} 
\maketitle

\begin{abstract}
Spatially correlated data with an excess of zeros, usually referred to as zero-inflated spatial data, arise in many disciplines. Examples include count data, for instance, abundance (or lack thereof) of animal species and disease counts, as well as semi-continuous data like observed precipitation. Spatial two-part models are a flexible class of models for such data. Fitting two-part models can be computationally expensive for large data due to high-dimensional dependent latent variables, costly matrix operations, and slow mixing Markov chains. We describe a flexible, computationally efficient approach for modeling large zero-inflated spatial data using the projection-based intrinsic conditional autoregression (PICAR) framework. We study our approach, which we call PICAR-Z, through extensive simulation studies and two environmental data sets. Our results suggest that PICAR-Z provides accurate predictions while remaining computationally efficient.  
 An important goal of our work is to allow researchers who are not experts in computation to easily build computationally efficient extensions to zero-inflated spatial models; this also allows for a more thorough exploration of modeling choices in two-part models than was previously possible. We show that PICAR-Z is easy to implement and extend in popular probabilistic programming languages such as \texttt{nimble} and \texttt{stan}. 
\end{abstract}

\noindent \textit{Keywords:} Zero-inflated spatial data, spatial statistics, two-part models, Bayesian analysis, basis representation, computational statistics

\newpage
\pagenumbering{arabic}  
\doublespacing

\section{Introduction}\label{Sec:Introduction}
Zero-inflated spatial data are spatially dependent observations characterized by an excess of zeros. Zero-inflated spatial data are common in many disciplines; for example, counts of harbor seals on glacial ice \citep{Hoef2007Seals}, annual mental health expenditures among US federal employees \citep{Neelon2011Expenditure}, and the number of torrential rainfall events in a region of interest \citep{lee2017applicability}. These data are typically a mixture of zeros and either discrete counts or positive real numbers. Standard probability distributions may not be appropriate for modeling zero-inflated data as they are unable to account effectively for the excess zeros \citep[cf.][]{Agarwal2002Zero, Rathbun2006Spatial,lambert1992zero}. 

A class of parametric mixture distributions called \textit{two-part models} \citep{Mullahy1986,lambert1992zero} are popular for modeling zero-inflated data. Two-part models account for both the excess zeros and the skewed distribution of non-zero values by using two latent random variables, one for occurrence and the other for prevalence. The occurrence process dictates whether a structural zero or non-zero value is observed, and the prevalence process determines the value of the structural non-zero observations. Two-part models have been extended to model zero-inflated spatial observations \citep[cf.][]{Agarwal2002Zero,Hoef2007Seals,Olsen2001Two} where the occurrence and prevalence random variables vary with respect to space (i.e. they are spatial random processes). Spatial two-part models are an extension of the well-known spatial generalized linear mixed models (SGLMMs) \citep{diggle1998model}, albeit with two separate latent spatial processes. 

Modeling large zero-inflated spatial datasets remains a key challenge from both a computational and modeling standpoint. First, fitting spatial two-part models can be computationally prohibitive for large datasets because these models are generally overparameterized \citep{recta2012two, Hoef2007Seals} with more unknown parameters than observations. For high-dimensional data, model-fitting can be computationally burdensome due to large matrix operations, estimation of high-dimensional latent random effects, and slow mixing Markov Chain Monte Carlo (MCMC) algorithms. Second, the underlying occurrence and prevalence spatial processes may potentially exhibit complex spatial dependence characteristics such as non-stationarity and anisotropy. Third, the occurrence and prevalence processes may be strongly correlated, and directly modeling the cross-correlations can be computationally expensive \citep{recta2012two}.

Novel modeling approaches for zero-inflated spatial data have been proposed, but these methods may not scale to larger datasets. In non-spatial settings, Frequentist methods have been used to fit two-part models using Gauss-Hermite quadrature \citep{min2005random}, expectation-maximization \citep{lambert1992zero,roeder1999modeling}, and restricted maximum quasi-likelihoods \citep{kim2012blup}. However, these methods do not address a key component of spatial two-part models - inference for the two $n$-dimensional vectors of latent spatial random effects from the occurrence and prevalence spatial processes. \citet{lyashevska2016mapping} employs a Monte Carlo maximum likelihood approach to model a moderately large zero-inflated spatial dataset ($n=4,029$), but this approach still requires estimation of the high-dimensional spatial random effects and costly matrix operations. 

In the Bayesian framework, the literature primarily focuses on the sophistication of zero-inflated spatial models. However, there is a dearth of studies examining the computational issues associated with large zero-inflated spatial datasets. Studies have modeled spatio-temporal dependence \citep{Fernandes2009ST, Neelon2016Spatiotemporal,Arcuti2016Bayesian}, addressed overdispersion \citep{Czado2008,lee2016spatial}, used skewed distributions \citep{Dreassi2014Small,liu2016analyzing}, used t-distributions to model heavy tailed behavior \citep{Neelon2015Semicontinuous}, and modeled prevalence with scale mixtures of normal distributions \citep{Fruhwirth2010} or Student-t processes \citep{bopp2020projecting}. Another study facilitates posterior sampling for zero-inflated negative binomial distributions (ZINB) by using latent variables that are represented as scale mixtures of normal distributions \citep{neelon2019bayesian}. A notable exception is \citet{Wang2014}, which models the presence and abundance of Atlantic cod in 1325 locations along the Gulf of Maine using predictive processes \citep{Banerjee2008Gaussian}. Predictive processes still requires $\mathcal{O}(nm^2+m^3)$ operations where $m$ is the number of knots and $n$ is the dataset size, and careful attention must be paid to knot selection \citep{guhaniyogi2011adaptive}.

In this study, we introduce a computationally efficient approach for fitting a broad range of two-part models to high-dimensional zero-inflated spatial data. We use projection-based intrinsic conditional autoregression (PICAR) \citep{lee2021picar} to reduce the dimensions of and correlation between the spatial random effects in two-part spatial models. The PICAR method represents the spatial random effects using empirical basis functions based on the Moran's I statistic and piecewise linear basis functions. Various basis representations have been directly or indirectly used to model spatial data; for instance, predictive processes \citep{Banerjee2008Gaussian}, random projections \citep{Guan_Haran_2018, guan2019fast, Banerjee2013, park2020reduced}, Moran's basis for areal models \citep{hughes2013dimension}, stochastic partial differential equations \citep{Lindgren2011}, kernel convolutions \citep{higdon1998process}, eigenvector spatial filtering \citep{griffith2003spatial}, and multi-resolution basis functions \citep{nychka2015multiresolution,katzfuss2017multi}, among others. \citet{hughes2013dimension} made a significant contribution in modeling large binary and count spatial datasets within the discrete spatial domain while de-confounding the spatial random effects; however, their methods does not extend to zero-inflated spatial data nor spatial datasets in the continuous domain.

To our knowledge, this is the first approach that readily lends itself to wide range of user-specified spatial two-part models while also reducing computational costs for large datasets. We demonstrate the applicability of our proposed approach (PICAR-Z) via simulation studies as well as two real-world applications - a bivalve species abundance dataset and ice thickness measurements in West Antarctica. Because our approach allows for efficient inference, it enables practitioners to thoroughly explore the advantages (and disadvantages) of various modeling choices in two-part models than was previously possible. 

The rest of the manuscript is as follows. In Section \ref{Sec:ZeroInflatedModels}, we introduce the general two-part modeling framework for zero-inflated data. We also provide an overview of spatial two-part models and examine the inherent modeling and computational challenges. In section \ref{Sec:PICARZ}, we propose a computationally efficient approach to fit high-dimensional spatial two-part models (PICAR-Z) and provide some implementation guidelines. We demonstrate the utility of PICAR-Z through four different simulation studies in Section \ref{Sec:SimulationStudy} as well as two high-dimensional environmental datasets in Section \ref{Sec:Applications}. Finally, a summary and directions for future research are provided in Section \ref{Sec:Discussion}. 

\section{Description of the Data}
Our proposed methodology (PICAR-Z) is motivated by two challenges in the environmental sciences (ecology and glaciology). We provide an overview of the motivating research questions as well as descriptions of the corresponding zero-inflated spatial datasets. 

\paragraph{Bivalve Species Abundance in the Wadden Sea:}
The Dutch Wadden Sea (DWS) is a UNESCO World Heritage region and a crucial protected ecological habitat comprised of sand barriers, salt marshes, mudflats, and gullies \citep{compton2013distinctly,lyashevska2016mapping}. Most notably, the DWS is a critical sanctuary for hundreds of thousands of shorebirds \citep{lyashevska2016mapping} as it serves as a vital feeding and breeding location for a wide array of bird species \citep{boere2012flyway}. The DWS plays a critical role in the bird species' yearly migratory journey by providing an abundance of food resources, such as the Baltic tellin (\textit{Macoma balthica}), a species of benthic invertebrates. The Baltic tellin is one of the most commonly found macrobenthic species in the western DWS \citep{van2003large} and one of the preferred preys of bird species \citep{lyashevska2016mapping}; however, there has been a steady decline in the population over the past two decades \citep{dairain2020high}. Recent studies have examined this decline in adult survival with respect to factors such as global warming \citep{beukema2020half}, habitat selection \citep{van2003large}, and proliferative disorders \citep{dairain2020high}. In addition, there are environmental factors that affect species survival such as the silt content, median grain size, and altitude \citep{lyashevska2016mapping}. Analyzing the abundance of Baltic tellin species, while considering spatial dependence and external environmental factors, can provide valuable insights for ecologists and conservationists. This information can guide resource allocation and the development of conservation policies, benefiting not only the Baltic tellin but also the diverse bird species within the Dutch Wadden Sea ecosystem.

We examine spatial abundance data of the Baltic tellin species from \citet{lyashevska2016mapping} originally obtained from the synoptic intertidal benthic surveys (SIBES) monitoring program \citep{compton2013distinctly,bijleveld2012designing}. The dataset includes counts of the Baltic tellin (\textit{Macoma balthica}) species sampled at $n=4,029$ locations. The dataset exhibits zero-inflation as $65.9\%$ of the locations have zero-counts as well as spatial dependence. The occurrence (presence vs. absence) and prevalence (values of positive counts) maps are provided in the supplement. Given the zero-inflated counts, spatial dependence, and large number of locations, computationally-efficient spatial two-part models are well-suited for analyzing this dataset.
\paragraph{Antarctic Ice Sheet Thickness:} Based on available geological records, mass loss from the Antarctic ice sheets can potentially lead to drastic global sea level rise \citep{deschamps2012ice}, in some cases up to 60 m \citep{fretwell2012bedmap2}. Existing studies \citep{serreze2011processes,zhang2007increasing} suggest that a portion of the Antarctic ice sheet could experience significant mass-loss in the next few centuries as a consequence of global climate change. This presents a significant threat since a substantial portion of the world's population resides in low-elevation coastal regions \citep{greve2011initial}. In fact, nearly eight percent of the global population is threatened by a mere five-meter rise in sea level \citep{nicholls2008global} and 13 percent of the global urban population is threatened by a ten-meter sea level rise \citep{mcgranahan2007rising}. Since Antarctic mass-loss is both a threat to heavily-populated coastal regions and a key indicator of global climate change, an important first step entails understanding the underlying spatial patterns of mass-loss by modeling the current thickness of the Antarctic ice sheet.  

We examine semicontinuous observations of ice thickness from the Bedmap2 dataset \citep{fretwell2012bedmap2}, generated using satellite altimetry, airborne and ground radar surveys, and seismic sounding. Similar to \citet{chang2016calibrating}, we examine gridded ice thickness observations at $20$ km resolution over a $171 \times 171$ grid spanning the entirety of Antarctica including vulnerable regions in the Amundsen Sea Embayment (West Antarctica). The resulting dataset consists of $n=29,241$ semi-continuous observations of ice sheet thickness where $10,327$ ($35.3\%$) are zeros. The observed dataset poses significant modeling and computational challenges due to: (1) high-dimensional observations; (2) the presence of zeros and positive thickness measurements; and (3) the data is collected at regularly-spaced intervals with many unobserved locations in the spatial domain. Traditional spatial generalized linear mixed models are unable to account for the zero-inflation and existing spatial two-part models are unable to scale to large datasets. In this Section~\ref{Subsec:Antarctic}, we utilize PICAR-Z to model the semicontinuous ice thickness and interpolate, or downscale, at unobserved locations. 

\section{Zero-inflated Spatial Models}\label{Sec:ZeroInflatedModels}
In this section, we provide an overview of the two-part modeling framework \citep{Mullahy1986,lambert1992zero} for spatially dependent zero-inflated observations \citep[cf.][]{Hoef2007Seals,Neelon2016Spatiotemporal,Wang2014}. Let $Z(\bs)$ be a zero-inflated observation for spatial location $\bs\subset\mathcal{D}$ within the spatial domain $\mathcal{D}\in\mathbb{R}^{2}$. In the spatial two-part modeling framework, $Z(\bs)$ are generated as follows. 
\begin{equation}\label{EQ:GenerateObs}
    Z(\bs) =
\left\{
	\begin{array}{ll}
		0  & \mbox{if } O(\bs) = 0 \\
		P(\bs) & \mbox{if } O(\bs) = 1.
	\end{array},
\right.
\end{equation}
where $O(\bs)$ and $P(\bs)$ are the spatial occurrence and prevalence processes, respectively. The occurrence process is typically specified as $O(\bs)\sim Bern(\cdot|\pi(\bs))$ with spatially varying probabilities $\pi(\bs)\in(0,1)$. The prevalence process is modeled as $P(\bs)\sim \tilde{F}(\cdot|\btheta(\bs))$ where $\tilde{F}(\cdot|\btheta(\bs))$ is a discrete or continuous probability distribution with spatially-varying parameters $\btheta(\bs)$. The key distinction from the univariate case is that the occurrence $O$ and prevalence $P$ random variables now vary across space; hence, we model the occurrence and prevalence as spatial random processes $O(\bs)$ and $P(\bs)$.

Spatial two-part models typically fall into two classes - hurdle and mixture models.  In \textit{hurdle models}, the occurrence random process $O(\bs)$ specifies which locations are associated with zero- or non-zero values. For the non-zero data, their respective positive values are generated by the prevalence random process $P(\bs)$. In the discrete case, $\tilde{F}(\cdot|\btheta(\bs))$ is a zero-truncated probability mass function such as the zero-truncated Poisson or the zero-truncated negative binomial distribution. For semi-continuous observations, $\tilde{F}(\cdot|\btheta(\bs))$ can be a probability density function with positive support such as a log-normal or gamma distribution. For \textit{mixture models}, the zero-valued observations can be generated by both processes $O(\bs)$ and $P(\bs)$. Here, $O(\bs)$ determines whether a location is classified as a structural zero or non-zero case. For the structural non-zero cases, the prevalence random process $P(\bs)$ generates both zeros and positive values. In the discrete case, $\tilde{F}(\cdot|\btheta(\bs))$ is a non-degenerate mass function such as the Poisson or Negative-Binomial distribution. For semi-continuous observations, $\tilde{F}(\cdot|\btheta(\bs))$ can be a censored model such as a Tobit Type I. 

\subsection{Modeling Framework: Spatial Two-Part Models}\label{SubSec:Two-PartGeneral}
Here, we outline the general hierarchical modeling framework for spatial two-part models. The occurrence process $O(\bs)$ corresponds to the Bernouilli probability distribution with either a probit or a logit link function. The prevalence process $P(\bs)$ employs the appropriate probability distribution based on the observation class (discrete vs. semi-continuous) and zero-inflation structure (hurdle vs. mixture). For hurdle models, a zero-truncated distribution (e.g., zero-truncated Poisson, zero-truncated negative binomial, lognormal, or gamma) is a sensible choice for $\tilde{F}(\cdot|\btheta)$. Mixture models utilize a distribution with non-negative support (e.g., Poisson, negative binomial, or Tobit Type I). For both processes, their respective linear predictors ($\bfeta_{o}$ and $\bfeta_{p}$) include the fixed and spatial random effects. The linear predictor for the occurrence process is $\bfeta_{o}=\bX\bbe_{o} + \bW_{o} + \bep_{o}$, where $\bbe_{o}$ and $\bW_{o}$ are the vectors of the fixed effects and spatial random effects, respectively, and $\bep_{o}$ are the iid observational errors. The linear predictor for the prevalence process $\bfeta_{p}$ can be constructed similarly to $\bfeta_{o}$. The latent spatial random sub-processes $W_o(\bs)$ and $W_p(\bs)$ can be modeled as a stationary zero-mean Gaussian process with a Mat\'ern covariance function \citep{stein2012interpolation}, a widely used class of stationary and isotropic covariance functions. Link functions $g_o(\cdot)$ and $g_p(\cdot)$ are specified according to the spatial processes. To complete the Bayesian hierarchical model, we designate prior distributions for the model parameters.  

The Bayesian hierarchical framework for spatial two-part models is:
\begin{align}\label{EQ:BayesTwoPartHier}
    \textbf{Data Model:}&\qquad \bZ|O(\bs),P(\bs) \sim F\big(\cdot|O(\bs),P(\bs)\big) \\
    \textbf{Process  Model:} & \qquad O(\bs)|\pi(\bs) \sim \mbox{Bern}(\cdot|\pi(\bs)) \nonumber \\
    & \qquad P(\bs)|\theta(\bs) \sim \tilde{F}(\cdot|\theta(\bs))\nonumber \\
\textbf{Sub-process  Model 1: } & \qquad \pi(\bs)|\eta_{o}(\bs)=g^{-1}_{o}(\eta_{o}(\bs))\nonumber\\
\textbf{(Occurrence) } & \qquad \eta_{o}(\bs)|\bbe_{o},W_{o}(\bs),\epsilon_{o}(s) =\bX(\bs)^{\prime}\bbe_{o} + W_{o}(\bs) +\epsilon_{o}(\bs) \nonumber\\
& \qquad \bW_{o}=(W_{o}(\bs_{1}),...,W_{o}(\bs_{n}))^{\prime} \nonumber\\
& \qquad \bW_{o}|\phi_{o},\sigma^{2}_{o}\sim \mathcal{N}(\pmb{0},\sigma^{2}_{o}\bR_{\phi_{o}}), \nonumber\\
& \qquad \epsilon(s)|\tau^{2}_{o}\sim \mathcal{N}(0,\tau^{2}_{o})\nonumber\\
\textbf{Sub-process  Model 2: } & \qquad \theta(\bs)|\eta_{p}(\bs)=g^{-1}_{p}(\eta_{p}(\bs)) \nonumber\\
\textbf{(Prevalence)}  & \qquad \eta_{p}(\bs)|\bbe_{p},W_{p}(\bs),\epsilon_{p}(\bs) =\bX(\bs)^{\prime}\bbe_{p} + W_{p}(\bs) +\epsilon_{p}(\bs) \nonumber\\
& \qquad \bW_{p}=(W_{p}(\bs_{1}),...,W_{p}(\bs_{n}))^{\prime} \nonumber\\
& \qquad \bW_{p}|\phi_{p},\sigma^{2}_{p}\sim \mathcal{N}(\bzero,\sigma^{2}_{p}\bR_{\phi_{p}}) \nonumber\\
& \qquad \epsilon(\bs)|\tau^{2}_{p}\sim \mathcal{N}(0,\tau^{2}_{p}) \nonumber\\
\textbf{Parameter  Model:} &\qquad  \bbe_{o}\sim p(\bbe_{o}),\quad \bbe_{p}\sim p(\bbe_{p}), \quad \phi_{o} \sim p(\phi_{o}) , \quad \phi_{p} \sim p(\phi_{p}) \nonumber\\
& \qquad \sigma^{2}_{o} \sim p(\sigma^{2}_{o}), \quad \sigma^{2}_{p} \sim p(\sigma^{2}_{p}), \quad \tau^{2}_{o} \sim p(\tau^{2}_{o}), \quad \tau^{2}_{p} \sim p(\tau^{2}_{p})\nonumber
\end{align}
where $F\big(\cdot|O(\bs),P(\bs)\big)$ is the distribution function of a spatial two-part model. From Equation \ref{EQ:BayesTwoPartHier}, the likelihood function $f\big(z|O(\bs),P(\bs)\big)$ is defined as:
\begin{equation}\label{EQ:Ch4TwoPartLikelihood}
    f\big(z|O(\bs),P(\bs)\big) =
\left\{
	\begin{array}{ll}
		\pi(\bs) +(1-\pi(\bs))\times \tilde{f}(0;\theta(\bs)),  & \mbox{if } z=0 \\
		(1-\pi(\bs))\times \tilde{f}(z;\theta(\bs)), & \mbox{if } z>0.
	\end{array},
\right.
\end{equation}
where $\pi(\bs)$ and $\theta(\bs)$ are the spatially-varying occurrence probabilities and prevalence intensities, respectively, and $\tilde{f}(z;\theta(\bs))$ is the density function of the prevalence process. Spatial two-part models fall into two classes (hurdle and mixture models). The key difference between both classes lies in the choice of $\tilde{F}(\cdot|\theta(\bs))$, the distribution of the prevalence process $P(s)$. We provide a detailed description of four spatial two-part models (count hurdle, semicontinuous hurdle, count mixture, and semicontinuous mixture) in the Supplement. 

\subsection{Computational Challenges}\label{SubSec:CompCost}
Spatial two-part models are subject to computational obstacles in the high-dimensional setting. Both the occurrence $O(\bs)$ and prevalence $P(\bs)$ processes include latent spatial random fields, which can be computationally prohibitive to model for even moderately large datasets (more than 1000 observations) \citep{haran2011gaussian}. These require a costly evaluation of an $n-$dimensional multivariate normal likelihood function with $\mathcal{O}(n^{3})$ operations at each iteration of the MCMC algorithm. The highly correlated spatial random effects can result in poor mixing Markov chains \citep[cf.][]{christensen2002bayesian,haran2003accelerating}. Another consideration is modeling the cross-covariance between the occurrence and prevalence processes, which also requires expensive matrix inversions and computing determinants \citep{recta2012two}.

Past studies employed Guass-Hermite quadrature \citep{min2005random}, expectation-maximization \citep{lambert1992zero,roeder1999modeling}, or restricted maximum quasi-likelihoods \citep{kim2012blup} to fit zero-inflated models. However, such approximations may not scale well with high-dimensional random effects \citep{neelon2016modeling} that exhibit spatial correlation. In the spatial setting, Monte Carlo Maximum Likelihood \citep{lyashevska2016mapping} and predictive processes \citep{Wang2014} have been incorporated into the two-part modeling framework. Yet, these approaches are still computationally costly and they may not scale well to larger datasets. For instance, \citet{lyashevska2016mapping} required over 72 hours to model $4029$ zero-inflated spatial observations. \citet{Wang2014} models a dataset with 1325 locations using predictive processes \citep{Banerjee2008Gaussian}, which scales at  $\mathcal{O}(nm^2+m^3)$ where $n$ is the number of observed locations and $m$ denotes the number of knot locations. Selecting the proper knot locations \citep{guhaniyogi2011adaptive} can also be challenging.

\section{A Computationally Efficient Approach for Fitting Two-Part Models}\label{Sec:PICARZ}
In this section, we propose a scalable method (PICAR-Z) for fitting high-dimensional spatial two-part models. Our approach builds upon the projection-based intrinsic conditional autoregression (PICAR) framework \citep{lee2021picar}. We present the general hierarchical modeling framework, practical guidelines for implementation, and a discussion of the computational speedup associated with PICAR-Z.  

Consider the vector of spatial random effects $\bW=(W(\bs_{1}),...,W(\bs_{n}))'$, which can be approximated as a linear combination of spatial basis functions:
$\mathbf{W}\approx \mathbf{\Phi}\bdel$ where $\mathbf{\Phi}$ is an $n\times p$ basis function matrix where each column denotes a basis function, and $\bdel\in\mathbb{R}^p$ are the re-parameterized spatial random effects (or basis coefficients). Moreover, $\bdel\sim \mathcal{N}(0,\Sigma_{\delta})$ where $\Sigma_\delta$ is the $p\times p$ covariance matrix for the coefficients. Basis functions can be interpreted as a set of distinct spatial patterns and a weighted sum of these patterns constructs a spatial random field. Basis representation has been a popular approach to model spatial data \citep[cf.][]{cressie2008fixed,Banerjee2008Gaussian,hughes2013dimension,Lindgren2011,rue2009approximate,haran2003accelerating,griffith2003spatial,higdon1998process,nychka2015multiresolution}. Basis representations tend to be computationally efficient \citep{cressie2015statistics} as they help bypass large matrix operations and reduce the dimensions of and correlation among the spatial random effects.
 
\subsection{Projection Intrinsic Autoregression (PICAR)}
The projection-based intrinsic conditional auto-regression (PICAR) approach consists of three components: (1) generate a triangular mesh on the spatial domain  $\mathcal{D}\subset \mathbb{R}^{2}$; (2) construct a spatial field on the mesh vertices using non-parametric basis functions; (3) interpolate onto the observation locations using piece-wise linear basis functions. Please see \citep{lee2021picar} for extensions to other classes of hierarchical spatial models such as spatial generalized linear mixed models, spatially-varying coefficients models, and ordinal spatial models.
 
\subsubsection*{Mesh Construction}
Prior to fitting the model, we generate a mesh enveloping the observed spatial locations via Delaunay Triangulation \citep{hjelle2006triangulations}. Next, the the spatial domain $D$ is partitioned into a collection of non-intersecting irregular triangles. The triangles can share a common edge, corner (i.e. nodes or vertices), or both. The mesh generates a latent undirected graph $G=\{V,E\}$, where $V =\{1,2, . . . , m\}$ are the mesh vertices and $E$ are the edges. Each edge $E$ is represented as a pair $(i, j)$ denoting the connection between $i$ and $j$. The graph $G$ is characterized by its weights matrix $\mathbf{N}$, an $m\times m$ matrix where $N_{ii}=0$ and $N_{ij} =1$ when mesh node $i$ is connected to node $j$ and $N_{ij} =0$ otherwise. The triangular mesh can built using the \textbf{R-INLA} package \citep{lindgren2015bayesian}. 

\subsubsection*{Moran's Basis Functions}\label{SubSec:Ch4Moran}
Next, we construct the Moran's basis functions \citep{hughes2013dimension, griffith2003spatial} on the set of mesh vertices $V$ of graph $G$. The Moran's basis functions refer to the leading $p$ eigenvectors of the Moran's operator $\mathbf{(I-11'/}m\mathbf{)N(I-11'/}m)$, where $\mathbf{I}$ is the identity matrix and $\mathbf{1}$ is a vector of $1$'s. Note that this operator is a component of the Moran's I statistic:
$$I(A)=\frac{m}{\mathbf{1'N1}}\frac{\mathbf{Z'(I-11'/}m)\mathbf{N(I-11'/}m\mathbf{)Z}}{\mathbf{Z'(I-11'/}m\mathbf{)Z}},$$ 
a diagnostic of spatial dependence \citep{moran1950notes} used for areal spatial data. Values of the Moran's I above $-\frac{1}{m-1}$ indicate positive spatial autocorrelation and values below $-\frac{1}{m-1}$ indicate negative spatial autocorrelation \citep{griffith2003spatial}. For the triangular mesh, the positive eigenvectors represent the patterns of spatial clustering, or dependence, among the mesh nodes, and their corresponding eigenvalues denote the magnitude of clustering. 
We construct the Moran's basis function matrix $\mathbf{M}\in \mathbb{R}^{m\times p}$, by selecting the first $p$ eigenvectors of the Moran's operator where $p\ll m$. Rank selection for $p$ proceeds via an automated heuristic \citep{lee2021picar} based on out-of-sample validation. A spatial random field can be constructed through a linear combinations of the Moran's basis functions (contained in matrix $\mathbf{M}$) and their corresponding weights $\delta\in \mathbb{R}^{p}$. 

\subsubsection*{Piece-wise Linear Basis Functions}
To complete the PICAR approach, we introduce a set of piece-wise linear basis functions \citep{brenner2007mathematical} to interpolate points within the triangular mesh (i.e. the undirected graph $G=(V,E)$). Following \citet{lee2021picar}, we construct a spatial random field on the mesh nodes $\mathbf{\tilde{W}}=(W(\bv_{1}),...,W(\bv_{m}))'$ where $\bv_{i}\in V$ and then project, or interpolate, onto the observed locations $\mathbf{W}=(W(\bs_{1}),...,W(\bs_{n}))'$ where $\bs_{i}\in \mathcal{D}$. The latent spatial random field $\mathbf{W}$ can be represented as $\mathbf{W=A\tilde{W}}$, where $\mathbf{A}$ is an $n\times m$ projector matrix containing the piece-wise linear basis functions. The rows of $\mathbf{A}$ correspond to an observation location $\bs_{i}\in\mathcal{D}$, and the columns correspond to a mesh node $\bv_{i}\in V$. The $i$th row of $\mathbf{A}$ contains the weights to linearly interpolate $W(s_{i})$. In practice, we use an $n\times m$ projector matrix $\mathbf{A}$ for fitting the hierarchical spatial model. For model validation and prediction, we generate an $n_{CV} \times m$ projector matrix $\mathbf{A}_{CV}$ that interpolates onto the $n_{CV}$ validation locations.

The PICAR approach \citep{lee2021picar} is specifically designed for spatial models in the continuous spatial domain and includes both a projection (P) and an intrinsic CAR (ICAR) component. The discretizing step of PICAR resembles models like ICAR for areal spatial data \citep{besag1995conditional}. However, because the full PICAR approach ultimately extends to the continuous domain and includes dimension-reduction, it allows practitioners to interpolate and quantify uncertainty at unobserved spatial locations on a continuous domain, as shown in examples in Section~\ref{Sec:Applications}. As discussed in \citet{lee2021picar} and \citet{hughes2013dimension}, $\bdel\sim \mcN(\bzero, (\bM'\bQ\bM)^{-1})$ where $\bQ$ represents the ICAR precision matrix. If $rank(\bM)=m$, then the latent process $\tilde{\bW}\sim \mcN(\bzero, \bQ^{-1})$, which is analogous to a Gaussian Markov random field with an ICAR precision matrix located on the mesh vertices. As for the projection component, PICAR represents the latent continuous spatial process as $\bW= \bA\tilde{\bW}= \bA\bM\bdel$ which is simply a projection of $\tilde{\bW}$ onto the continuous domain through projection matrix $\bA$.

PICAR’s projected Moran’s basis functions ($\bA\bM$) and the restricted Moran’s basis functions (HH) from \citet{hughes2013dimension} share some features. Both bases are eigenvectors of a variant of the Moran’s operator, $\mathbf{(I-11'/}m\mathbf{)N(I-11'/}m)$ for PICAR and $\mathbf{(I-\bP)N(I-\bP)}$ for HH where $\bP=\bX(\bX'\bX)^{-1}\bX'$. However, PICAR’s bases can project onto the continuous spatial domain using projector matrix $\bA$, while the HH bases are limited to the discrete spatial domain. Recent critiques of low-rank restricted spatial regression models (RSR) \citep{zimmerman2022deconfounding,khan2022restricted} show that RSR approaches like HH have poor inferential and predictive performance compared to non-RSR SGLMMs and even non-spatial models. However, PICAR and PICAR-Z do not suffer from these issues because neither is an RSR approach; the focus is entirely on creating a computationally efficient approach and no attempt is made to examine spatial confounding.

\subsection{PICAR Approach for Zero-inflated Spatial Data}
The PICAR approach readily extends to spatial two-part models (Equation \ref{EQ:BayesTwoPartHier}). Using PICAR, we approximate both the latent spatial occurrence $O(\bs)$ and prevalence $P(\bs)$ processes as an expansion of Moran's basis functions \citep{griffith2003spatial, hughes2013dimension}. PICAR-Z approximates the spatial random effects for the occurrence and prevalences processes, respectively, as $\bW_{o}\approx \bA_{o}\bM_{o}\bdel_{o}$ and $\bW_{p}\approx \bA_{p}\bM_{p}\bdel_{p}$ using projector matrices $\bA_{o}$ and $\bA_{p}$, Moran's basis functions matrices $\bM_{o}$ and $\bM_{p}$, and basis coefficients $\bdel_{o}$ and $\bdel_{p}$. The PICAR-Z hierarchical framework for spatial two-part models is:
\begin{align*}
\textbf{Data Model: } & \qquad Z(\bs)|O(\bs),P(\bs) \sim F\big(\cdot|O(\bs),P(\bs)\big)\\
\textbf{Process Model: } & \qquad O(\bs)|\pi(\bs) \sim \mbox{Bern}(\pi(\bs))\\
& \qquad P(\bs)|\theta(\bs) \sim \tilde{F}(\cdot | \theta(\bs))\\
\textbf{Sub-process  Model 1: }& \qquad \pi(\bs)|\eta_{o}(\bs)=g^{-1}_{o}(\eta_{o}(\bs))\\
\textbf{(Occurrence) } & \qquad \eta_{o}(\bs)|\bbe_{o},\bdel_{o}= \bX(s)^{\prime}\bbe_{o} + [\bA_{o}\bM_{o}\bdel_{o}](\bs)\\
& \qquad \bdel_{o}|\tau_{o} \sim \mathcal{N}(\bzero,\tau_{o}^{-1}(\bM_{o}'\bQ_{o}\bM_{o})^{-1})\\
\textbf{Sub-process  Model 2: } & \qquad \theta(\bs)|\eta_{p}(\bs)=g^{-1}_{p}(\eta_{p}(\bs))\\
\textbf{(Prevalence)} & \qquad  \eta_{p}(\bs)|\bbe_{p},\bdel_{p}=\bX(s)^{\prime}\bbe_{p} + [\bA_{p}\bM_{p}\bdel_{p}](\bs) \\
& \qquad \bdel_{p}|\tau_{p}\sim \mathcal{N}(\bzero,\tau_{p}^{-1}(\bM_{p}'\bQ_{p}\bM_{p})^{-1}),\\
\textbf{Parameter  Model: } & \qquad \bbe_{o}\sim \mcN(\bmu_{\beta_o}, \Sigma_{\beta_o}),\quad \bbe_{p}\sim \mcN(\bmu_{\beta_p}, \Sigma_{\beta_p})\\
& \qquad \tau_o \sim \mcG(\alpha_{\tau_o},\beta_{\tau_o}), \quad \tau_p \sim \mcG(\alpha_{\tau_p},\beta_{\tau_p})
\end{align*}
where $F\big(\cdot|O(\bs),P(\bs)\big)$ is the distribution of the specified two-part model such as the hurdle and mixture models. $\tilde{F}(\cdot|\theta(\bs))$ denote the distribution function of the prevalence process. $\bA_{o}$ and $\bA_{p}$ are the projectors matrices for the occurrence and prevalence processes, respectively. For the occurrence and prevalence processes, we now incorporate the Moran's basis functions matrices $\mathbf{M}_{o}$ and $\mathbf{M}_{p}$, the basis coefficients $\bdel_{o}$ and $\bdel_{p}$, and the precision parameters $\tau_{o}$ and $\tau_{p}$. $\bQ$, the $m \times m$ prior precision matrix for the mesh vertices, is typically fixed prior to model fitting (see \citet{lee2021picar} for additional details). $[\bA_{o}\bM_{o}\bdel_{o}](\bs)$ denotes the value of the basis expansion $\bA_{o}\bM_{o}\bdel_{o}$ corresponding to location $\bs$. The interpretation of $[\bA_{p}\bM_{p}\bdel_{p}]$ follows similarly. 

The PICAR-Z approach is amenable to be modified to capture the cross-covariance between the occurrence $O(\bs)$ and prevalence $P(\bs)$ processes. Past studies have examined methodology for estimating the cross-covariance between two spatial random fields \citep{oliver2003gaussian, recta2012two}. We extend the general modeling framework from \citet{recta2012two} to high-dimensional settings by imposing correlation on the dimension-reduced Moran's basis coefficients $\bdel_{o}$ and $\bdel_{p}$. We provide additional details for modeling the cross-correlation under the PICAR-Z framework in the supplement. 

\subsection{Tuning Mechanisms}\label{SubSec:TuningMech}
Though most of the PICAR-Z approach is readily automated, there are key tuning mechanisms left to the practitioner: (1) selecting the rank of Moran's basis functions matrices ($p_o$ and $p_p$); (2) specifying the precision matrices of the mesh vertices ($\bQ_o$ and $\bQ_p$); and (3) identifying the appropriate two-part model. Here, we examine these tuning mechanisms in detail and provide practical guidelines for implementation. 

First, we provide an automated heuristic to select the appropriate ranks ($p_{o}$ and $p_{p}$) of the Moran's basis function matrices for both processes, $\mathbf{M}_{o}$ and $\mathbf{M}_{p}$.
First, we generate two augmented datasets, $\mathbf{Z}_{o}^{*}$ and $\mathbf{Z}_{p}^{*}$, constructed from the original zero-inflated spatial dataset $\mathbf{Z}$. The first dataset is generated as: 
\begin{equation}
    Z_{o}^{*}(s) =
\left\{
	\begin{array}{ll}
	    0,  & \mbox{if } Z(s)=0 \\
		1, & \mbox{if } Z(s)>0.
	\end{array},
\right.
\end{equation}
The second dataset $\mathbf{Z}_{p}^{*}\in \mathbb{R}^{n_{p}}$ is the collection of all observations such that $Z(s)>0$ and $n_{p}$ corresponds to the sample size of $\mathbf{Z}_{p}^{*}$. Next, we generate a set $\mathcal{P}$ consisting of $h$ equally spaced points within the interval $[2,P]$ where $P$ is the maximum rank and $h$ is the interval resolution ($h=P-1$ by default). Here, $P<m$ and both $P$ and $h$ are chosen by the user. 

For the augmented dataset $\mathbf{Z}_{o}^{*}$, we proceed in the following way. For each $p\in\mathcal{P}$, we construct an $n\times(k+p)$ matrix of augmented covariates $\tilde{X}_{o}=[X \quad \mathbf{A_{o}M}_{p}]$ where $X\in \mathbb{R}^{n\times k}$ is the original covariate matrix, $\mathbf{A}_{o} \in \mathbb{R}^{n\times m}$ is the projector matrix, and $\mathbf{M}_{p} \in \mathbb{R}^{m\times p}$ are the leading $p$ eigenvectors of the Moran's operator. Next, we use maximum likelihood approaches to fit the appropriate generalized linear model (GLM) for binary responses with a logit or probit link function. Finally, we set $p_{o}$ to be the rank $p$ that yields the lowest out-of-sample root mean squared prediction error (rmspe) or area under the ROC curve (AUC). 

We implement a similar procedure for the second augmented dataset $\mathbf{Z}_{p}^{*}\in\mathbb{R}^{n_{p}}$. For each $p\in\mathcal{P}$, we construct an $n_{p}\times(k+p)$ matrix of augmented covariates $\tilde{\bX}_{p}=\begin{bmatrix}
\bX_{p}& \mathbf{A}_{p}\bM_{p}
\end{bmatrix}$ where $\bX_{p}\in \mathbb{R}^{n_{p}\times k}$ is the matrix of covariates, $\mathbf{A}_{p} \in \mathbb{R}^{n_{p}\times m}$ is the projector matrix, and $\mathbf{M_{p}} \in \mathbb{R}^{m\times p}$ are the leading $p$ eigenvectors of the Moran's operator. Note that the rows of $\bX_{p}$, $\mathbf{A}_{p}$, and $\mathbf{M_{p}}$ correspond to the $n_{p}$ locations with non-zero values. Next, we use maximum likelihood approaches to fit the appropriate generalized linear model (GLM) for positive responses. For count data, the likelihood function is a zero-truncated Poisson distribution. For semi-continuous data in the hurdle model framework, we employ a lognormal distribution as the likelihood function. For semi-continuous data in the mixture model framework, we simply fit the traditional linear model. Then, we set $p_{p}$ to be the rank $p$ that yields the lowest out-of-sample root mean squared prediction error (rmspe).

Next, we provide some choices for $\mathbf{Q}$, the prior precision matrix for the mesh vertices $\mathbf{\tilde{W}}$. By default, we set $\mathbf{Q}$ to be the precision matrix of an intrinsic conditional auto-regressive model (ICAR). Another option would set $\mathbf{Q}$ as the precision matrix of a conditional auto-regressive model (CAR) with estimable autocorrelation parameter $\rho\in(0,1)$ such that $\mathbf{Q}=(\mathbf{N1}-\rho \mathbf{N})$, where $\mathbf{N}$ is the adjacency matrix. In practice, estimating $\rho$ could potentially offset the computational gains of the PICAR-Z approach. Note that there are two latent processes (occurrence and prevalence), hence we must estimate two additional parameters, $\rho_o$ and $\rho_p$, and construct two precision matrices $\bQ_{O}$ and $\bQ_{p}$. At each iteration of the MCMC algorithm, both $\rho_o$ and $\rho_p$ must be sampled, and determinants $|\bQ_{O}|$ and $|\bQ_{p}|$ must be recomputed. This would increase computational costs on the order of $\mathcal{O}(\frac{2}{3}m^{3})$ operations (i.e. two additional Cholesky decompositions of an $m\times m$ precision matrix). Another alternative is setting $\mathbf{Q}=I$, where the basis coefficients $\bdel_o$ and $\bdel_p$ are uncorrelated a priori.

Specifying the class (hurdle vs. mixture) of two-part model is an active area of research \citep[cf.][]{feng2021comparison,vuong1989likelihood,neelon2016modeling}. Past studies have selected the appropriate model class using the Akaike Information Criterion (AIC) \citep{feng2021comparison} or the Vuong test statistics \citep{vuong1989likelihood}. However, the choice between a hurdle and mixture model depends on the aims of the investigator and prior scientific knowledge regarding the zero-generating processes (i.e. should the prevalence process also generate zeros). For practitioners, we suggest conducting a sensitivity analysis using both mixture and hurdle models, and then select the appropriate model based on out-of-sample validation. Since PICAR-Z is computationally efficient and scales to larger datasets, conducting such a sensitivity analysis should be feasible in many settings. 

\subsection{Computational Advantages}
The PICAR-Z approach requires shorter walltimes per iteration (of the MCMC algorithm) as well as fewer iterations for the Markov chain to converge. The computational speedup results from exploiting lower-dimensional and weakly correlated basis coefficients $\bdel_{o}$ and $\bdel_{p}$ and also bypassing expensive matrix operations (e.g. Cholesky decompositions). The PICAR-Z approach has a computational complexity of $\mathcal{O}(2np)$ as opposed to $\mathcal{O}(2n^{3})$ for the full hierarchical spatial two-part model. 

We examine mixing in MCMC algorithms within the context of spatial two-part models. The PICAR-Z approach generates a faster mixing MCMC algorithm than fitting the full two-part model using the `reparameterized' method \citep{christensen2002bayesian} and a competing approach using bisquare basis functions \citep{cressie2008fixed}. This is corroborated by the larger effective sample size per second (ES/sec), which approximates the rate at which samples are produced from an MCMC algorithm that are equivalent to samples from an IID sampler. Larger values of ES/sec indicates faster mixing. In the simulated examples (Section \ref{Sec:SimulationStudy}), the PICAR-Z approach returns a larger ES/sec than the `reparameterized' approach across all model parameters and spatial random effects (see supplement). 

For PICAR, the two major computational bottlenecks are constructing the Moran's operator and computing its $k$-leading eigencomponents. The Moran's operator requires the matrix operation $\mathbf{(I-11'/}m)\mathbf{N}(\mathbf{I-11'}/m)$ which results in $2m^{3}-m^{2}$ floating point operations (FLOPs) where $m$ is the number of mesh vertices. For dense mesh structures (large $m$), we can generate the Moran's operator by leveraging the embarrassingly parallel operations as well the sparsity of the weights matrix $\mathbf{N}$. Next, the first $k$ eigencomponents of the Moran's Operator can be computed using a partial eigendecomposition approach such as the Implicitly Restarted Arnoldi Method \citep{lehoucq1998arpack} from the \textbf{RSpectra} package \citep{RSpectra2019}. Since the PICAR-Z approach generally selects the leading $p_o$ or $p_p$ eigenvectors where $p_o\ll m$ and $p_p\ll m$, an expensive full eigendecomposition of the Moran's operator is not necessary. 

\section{Simulation Study}\label{Sec:SimulationStudy}
In this section, we conduct an extensive simulation study focusing on four commonly-used spatial two-part models: (1) hurdle model with count data; (2) mixture model with count data; (3) hurdle model with semi-continuous data; and (4) mixture model with semi-continuous data. In addition, we provide comparisons to a low-rank approach with nested bisquare basis functions \citep{sengupta2013hierarchical, cressie2008fixed} and a `reparameterized' method \citep{christensen2002bayesian}. 

\subsection{Simulation Study Design}
For all four two-part models, we simulate $B=100$ samples for a total of $4\times 100=400$ datasets in the simulation study. In each sample, we generate a set of $1400$ randomly-selected locations on the unit square. $1000$ observations are allocated for model-fitting and the remaining $400$ reserved for model validation. We chose a smaller sample size $n=1000$ to allow for comparisons against a `reparameterized' method, which may be difficult to implement with larger datasets (see Section \ref{SubSec:CompCost}). Please see the supplement for details on the `reparameterized' approach.  

For each sample, we randomly generate a matrix of covariates $\bX=[\bX_1 , \bX_2]$. We use the same regression coefficients $\bbe_{o}=\bbe_{p}=(1,1)^{T}$ for all datasets in the simulation study. The spatial random effects $\bW_o$ and $\bW_p$ are generated from the multivariate Gaussian process proposed in \citet{recta2012two}. Note that the prohibitively high computational costs made it challenging to explore model structures carefully and extend them to higher-dimensional settings. 
\begin{equation*}
    \begin{bmatrix}
\bW_o\\
\bW_p
\end{bmatrix} \sim \mcN \Bigg(
    \begin{bmatrix}
\bzero\\
\bzero
\end{bmatrix}, 
    \begin{bmatrix}
\bC(\cdot|\nu_o,\phi_o,\sigma^2_o) & \rho\bL_o\bL_p^{T}\\
\rho\bL_o\bL_p^{T}& \bC(\cdot|\nu_p,\phi_p,\sigma^2_p )
\end{bmatrix}
\Bigg).
\end{equation*}
where $\bC(\cdot|\nu_o,\phi_o,\sigma^2_o)$ and $\bC(\cdot|\nu_p,\phi_p,\sigma^2_p )$ are covariance matrices for $\bW_o$ and $\bW_p$, respectively. $\rho$ represents cross-correlation between the occurrence and prevalence processes at the same location. We fix $\rho=0.7$ to impose moderate positive cross-correlation between the occurrence and prevalence processes. $\bL_o$ and $\bL_p$ are the lower-triangular Choleski factors of $\bC(\cdot|\nu_o,\phi_o,\sigma^2_o)$ and $\bC(\cdot|\nu_p,\phi_p,\sigma^2_p )$, respectively. That is, $\bC(\cdot|\nu_o,\phi_o,\sigma^2_o )=\bL_o\bL_o^{T}$ and $\bC(\cdot|\nu_p,\phi_p,\sigma^2_p )=\bL_p\bL_p^{T}$. The covariance matrices, $\bC(\cdot|\nu_o,\phi_o,\sigma^2_o)$ and $\bC(\cdot|\nu_p,\phi_p,\sigma^2_p )$ are from the Mat\'ern class \citep[cf.][]{williams2006gaussian, rasmussen2004gaussian} of covariance functions with parameters $\nu_o=\nu_p=0.5$, $\sigma^{2}_o=\sigma^{2}_p=1$, and $\phi_o=\phi_p=0.2$. We use the exponential covariance function ($\nu=0.5$) to generate a ``rough'' latent spatial process that is not mean square differentiable \citep{rasmussen2004gaussian}. 

We first generate realizations from the latent occurrence process $O(\bs)$ such that the underlying probability surface is modeled as $\pi(\bs)=\mbox{logit}^{-1}(\bX(\bs)^{\prime}\bbe_{O}+\bW_O(\bs))$. Next, we generate realizations from the prevalence process $P(\bs)$ from the corresponding prevalence distribution $\tilde{F}(\cdot|\theta(\bs))$ (Equation \ref{EQ:BayesTwoPartHier}) with spatially varying intensity (or mean) processes $\theta(\bs)$. We use a zero-truncated Poisson, Lognormal, Poisson, and Type I Tobit distribution for the hurdle count, hurdle semi-continuous, mixture count, and mixture semi-continuous cases, respectively. For the hurdle count and mixture count cases, we model the underlying intensity process as $\theta(\bs)=\exp\{\bX(\bs)^{\prime}\bbe_P + \bW_p(\bs)\}$. For the semi-continuous models, we specify $\theta(\bs)=(\bmu(\bs),\tau^2)$ with the mean process $\bmu(\bs)=\bX(\bs)^{\prime}\bbe_P + \bW_p(\bs)$ and nugget variance $\tau^2=0.1$. Finally, the observed data are drawn from the respective distribution of the spatial two-part model $F(\cdot|O(\bs),P(\bs))$. Our simulated datasets exhibit balance in the number of zero- and non-zero-valued observations (see summary in the supplement)

\subsubsection*{Implementation and Competing Methods}
To complete the hierarchical framework, we specify prior distributions for the zero-inflated spatial model parameters. We assign a multivariate normal prior for the regression coefficients where $\bbe_O\sim \mcN(\bzero, 100\mcI)$ and $\bbe_P\sim \mcN(\bzero, 100\mcI)$. For the variance of the spatial basis coefficients, we specify a non-informative inverse gamma priors where $\sigma^{2}_{O}\sim \mathcal{IG}(0.002, 0.002)$ and $\sigma^{2}_{P}\sim \mathcal{IG}(0.002, 0.002)$. The cross-correlation coefficient between the occurrence and prevalence processes $\rho$ follows a uniform distribution $\rho \sim \mathcal{U}(-1,1)$. For the semi-continuous cases, the nugget variance $\tau^2_\epsilon$ for the prevalence process follows an inverse gamma distribution $\tau^2_\epsilon\sim \mathcal{IG}(0.002, 0.002)$. For the PICAR-Z approach, we ran $150,000$ iterations of the MCMC algorithm. The MCMC algorithm is implemented using the programming language \texttt{nimble} \citep{nimble2017}. The selected rank, $p_o$ and $p_p$, varies across datasets and the class of two-part models. The median rank for the occurrence processes ($p_o$), is smaller than the median rank for the prevalence processes ($p_p$), which is consistent with results from a previous study \citep{lee2021picar}. We provide a table and summary of the chosen ranks in the supplement.

We compare the PICAR-Z method against the low-rank (bisquare) approach \citep{sengupta2013hierarchical, cressie2008fixed} as well as the `reparameterized' approach \citep{christensen2002bayesian}. Due to computational constraints, we elected to use the approach from \citet{christensen2002bayesian} over the full spatial hierarchical two-part model from Section \ref{SubSec:Two-PartGeneral}. The `reparameterized' approach is, by design, a computationally efficient method for modeling latent spatial random processes as it improves mixing in the MCMC algorithm by considerably reducing the correlation in the spatial random effects. This method preserves the rank of the spatial random effects and improves mixing in the MCMC algorithm by considerably reducing the correlation in the spatial random effects. For the `reparameterized' approach, we assume that the class of covariance function (Mat\'ern) and the smoothness parameter ($\nu=0.5$) are known a priori, which may not necessarily be the case in other scenarios. Additional details for the competing approaches are provided in the supplement. 

We provide the following validation metrics averaged over all samples in the simulation study: (1) out-of-sample root mean squared prediction error (rmspe total) for all observations; (2) area under the receiver operating characteristic curve (AUC) for the zero-valued observations; and (3) rmspe for the non-zero-valued observations (rmspe positive). The AUC is used to assess how well each approach classifies zero-valued observations in a binary classification setting. The out-of-sample root mean squared prediction error (rmspe) is $\mbox{rmspe}=\sqrt{\frac{1}{n_{CV}}\sum_{i=1}^{n_{CV}}(Y^{*}_{i}-\hat{Y}^{*}_{i})^{2}}$, where $n_{CV}=400$, $Y^{*}_{i}$'s denote the i-th value in the validation sample, and $\hat{Y}^{*}_{i}$'s are the predicted values at the $i$-th location. In addition, we compare the computational walltimes to draw $150,000$ samples from the respective posterior distributions via MCMC. We asses convergence of the Markov chains using the batch means standard errors. The computation times are based on a single 2.4 GHz Intel Xeon Gold 6240R processor. All the code was run on the George Mason University Office of Research Computing (ORC) HOPPER high-performance computing infrastructure.

\subsection{Results}\label{SubSec:SimExamplesTwoPart}
\begin{figure}[ht]
    \centering
    \includegraphics[width=1\textwidth]{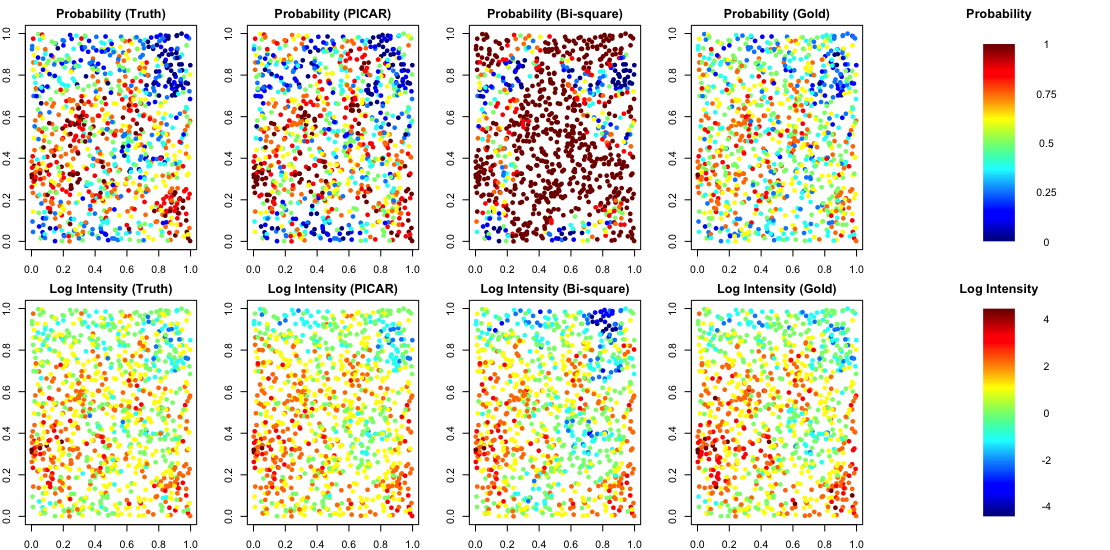}
    \caption{Prediction results from a single simulated example. Data are generated from the spatial count mixture model in Section \ref{Sec:SimulationStudy}. Top row includes the true and predicted probability surfaces $\pi(\bs)$ for the occurrence random process $O(\bs)$. Bottom row presents the true and predicted log-intensity surfaces $\log(\theta(\bs))$ for the prevalence random process $P(\bs)$. Column 1 presents the true latent probability and log-intensity surfaces. Columns 2-3 include the predicted surfaces for PICAR-Z (column 2), low-rank approach with bisquare basis functions (column 3), and the `reparameterized' approach (column 4). In the fifth column, a color scale is provided for the probability and log-intensity surfaces.}\label{Fig:SimStudy}
\end{figure}

Table \ref{Tab:SimulationStudy} contains the out-of-sample prediction results for the entire validation sample (rmspe), positive-valued observations (rmspe), and zero vs. non-zero values (AUC) as well as the average model-fitting walltimes. Results of the simulation study suggest that PICAR-Z outperforms both competing approaches in prediction across all four classes of two-part models (see Table \ref{Tab:SimulationStudy}). All approaches perform comparably for binary classification of the zero vs. non-zero cases, as corroborated by similar AUC values. However, the PICAR-Z methods (with and without cross-correlation) provide more accurate predictions for the non-zero (i.e positive-valued) observations, in comparison to the other two methods. Estimating the correlation parameter does not strongly affect accuracy, save for the semi-continuous hurdle case. Note that the PICAR-Z approach outperformed the `reparameterized' approach in predictive performance, which is consistent with results from past studies that examined basis representations of spatial latent fields \citep{bradley2019bayesian, lee2021picar}. Figure \ref{Fig:SimStudy} provides a visual representation of the latent probability $\pi(\bs)$ and log-intensity $log(\theta(\bs))$ surfaces. 

We also consider mixing in MCMC algorithms by examining the effective sample size per second (ES/sec), or the rate at which independent samples are generated by the MCMC algorithm. Larger values of ES/sec are indicative of faster mixing Markov chains. Note that PICAR-Z approach generates a faster mixing MCMC algorithm than the `reparameterized' approach \citep{christensen2002bayesian}, a method specifically designed to improve mixing for SGLMMs. For model parameters $\beta_{1o}$, $\beta_{2o}$, $\beta_{1p}$, and $\beta_{1p}$, PICAR-Z yields an ES/sec of $218.89$, $214.15$, $44.00$, and $43.83$, respectively. The `reparameterized' approach returns an ES/sec $0.66$, $0.63$, $0.26$, and $0.25$, respectively. For the spatial random effects $\mathbf{W}_{o}(s)$ and $\mathbf{W}_{p}(s)$, the median ES/sec is $53.09$ and $28.70$ for the PICAR-Z approach and $0.71$ and $0.40$ for the `reparameterized' approach, an improvement by a factor of roughly $74.3$ and $72.5$. Across all four model classes, the PICAR-Z approach has shorter walltimes to run $150,000$ iterations of the Metropolis-Hastings algorithm than low-rank (bisquare) and the `reparameterized' approach (Table \ref{Tab:SimulationStudy}). Against the `reparameterized' approach, PICAR-Z exhibits a speed-up factor of roughly $152.4$, $121.2$, $203.9$, and $177.4$ for the count hurdle, semi-continuous hurdle, count mixture, and semi-continuous mixture models, respectively. We also conducted a sensitivity analysis regarding the proportion of zeros within the sample and various model performance metrics. Results (see supplement) indicate that datasets with a low proportion of zeros have lower AUC (poor classification) than datasets with larger proportion of zeros. Low proportions of zeros are linked with shorter model-fitting walltimes. Boxplots of the relevant metrics - Total RMSPE, Non-Zero RMPSE, AUC for the zero-valued observations, and walltimes - are also provided in the supplement.

\begin{table}[ht]
\centering
\begin{tabular}{|c|c|cccc|}
  \hline
  Two-Part& Fitting & Total & Non-Zero & Zero  & Time \\ 
Model& Method & RMSPE & RMSPE & AUC & (mins) \\ 
  \hline
 \multirow{2}{*}{Count}& PICAR-Z & 2.182 & 2.594 & 0.726 & 1.500 \\ 
 \multirow{2}{*}{Hurdle} &PICAR-Z (Cor) & 2.195 & 2.581 & 0.725 & 2.100 \\ 
  & Bisquare & 2.252 & 2.701 & 0.722 & 2.200 \\ 
  & Reparameterized & 2.209 & 2.952 & 0.723 & 228.600 \\ 
  \hline
  \multirow{2}{*}{Semi-continuous} & PICAR-Z & 2.376 & 2.917 & 0.737 & 1.900 \\ 
  \multirow{2}{*}{ Hurdle} & PICAR-Z (Cor) & 2.353 & 2.953 & 0.736 & 3.500 \\ 
  & Bisquare & 2.834 & 3.503 & 0.731 & 2.200 \\ 
  & Reparameterized & 2.532 & 3.285 & 0.734 & 230.300 \\ 
  \hline
  \multirow{2}{*}{Count} & PICAR-Z & 2.366 & 3.374 & 0.805 & 1.900 \\ 
  \multirow{2}{*}{Mixture} & PICAR-Z (Cor) & 2.366 & 3.364 & 0.806 & 2.600 \\ 
  & Bisquare & 2.528 & 3.462 & 0.802 & 2.700 \\ 
  & Reparameterized & 2.428 & 3.704 & 0.806 & 387.400 \\ 
  \hline
  \multirow{2}{*}{Semi-continuous}  &PICAR-Z & 0.515 & 0.825 & 0.854 & 2.200 \\ 
  \multirow{2}{*}{Mixture} & PICAR-Z (Cor) & 0.518 & 0.831 & 0.859 & 3.200 \\ 
  & Bisquare & 0.563 & 0.882 & 0.789 & 2.700 \\ 
  & Reparameterized & 0.676 & 1.081 & 0.734 & 390.200 \\ 
   \hline
\end{tabular}
\caption{Simulation Study Results: Median values for all 100 samples in the simulation study. Results are grouped by two-part model (hurdle vs. mixture), data type (counts vs. semi-continuous), and approach (PICAR-Z, PICAR-Z with cross-correlation, low-rank (bisquare), and the `reparameterized' approach). Results include the root mean squared prediction error (rmspe) for the entire validation dataset and the non-zero data.  For zero- vs. non-zero classification, we report the area under the ROC curve (AUC). Model-fitting walltimes are reported in minutes.} \label{Tab:SimulationStudy}
\end{table}

\section{Applications}\label{Sec:Applications}
We demonstrate the scalability and flexibility of PICAR-Z on two large environmental datasets with spatially-referenced zero-inflated observations - zero-inflated counts of a bivalve species and high-resolution ice thickness measurements over West Antarctica. 

\subsection{Bivalve Species in the Dutch Wadden Sea}\label{Subsec:WaddenSea}
We randomly select $3,220$ observations to fit our model and hold out $806$ observations for validation. Covariates include environmental variables that affect the abundance of the \textit{Macoma balthica} species such as: (1) median grain size of the sediments; (2) silt content of the sediments; and (3) altitude. Using PICAR-Z, we fit the hurdle count model and the zero-inflated Poisson model (mixture). We construct a triangular mesh with $m=4,028$ mesh vertices. In both the count hurdle and mixture cases, the automated heuristic (Section \ref{SubSec:TuningMech}) chose ranks $p_{o}=14$ and $p_{p}=64$ for the occurrence and prevalence processes, respectively. 

We employ similar model specifications and prior distributions as in the simulated examples (Section \ref{SubSec:SimExamplesTwoPart}), including a comparisons to the low-rank (bisquare) approach. Comparative results are provided in the Table ~\ref{Table:MacomaResults}. Both the PICAR-Z and correlated PICAR-Z approach outperforms low-rank approaches in predictions and with shorter model-fitting walltimes. These results hold in both the hurdle and mixture modeling approaches. Note that PICAR-Z provides more accurate predictions, compared to the low-rank (bisquare) approach, among the non-zero observations. Comparisons to the `reparameterized' approach \citep{christensen2002bayesian} are computationally prohibitive due to the long wall times associated with the MCMC algorithms. Under PICAR-Z, both the hurdle and mixture models provide comparable out-of-sample predictions, yet the hurdle model has a shorter walltime; therefore, we recommend the count hurdle model for this particular case. We present the inferential results for the regression coefficients $\bbe_{o}$ and $\bbe_{p}$ in the supplement. 

\begin{table}[ht]
\centering
\begin{tabular}{|c|c|cccc|}
  \hline
  Two-Part& Fitting & Total & Non-Zero & Zero  & Time \\ 
Model& Method & RMSPE & RMSPE & AUC & (mins) \\ 
  \hline
\multirow{2}{*}{Count} &  PICAR-Z  & 4.50 & 7.41 & 0.64 & 2.53 \\ 
\multirow{2}{*}{Hurdle} &    PICAR-Z (Cor) & 4.50 & 7.40 & 0.64 & 3.05 \\ 
 & Bisquare & 4.80 & 7.83 & 0.64 & 3.15 \\ 
   \hline
\multirow{2}{*}{Count}  &   PICAR-Z  & 4.51 & 7.41 & 0.64 & 4.53 \\ 
\multirow{2}{*}{Mixture} & PICAR-Z (Cor) & 4.52 & 7.41 & 0.64 & 4.88 \\ 
 & Bisquare & 4.82 & 7.89 & 0.64 & 4.95 \\ 
   \hline
\end{tabular}
\caption{Bivalve Species Results: Results are grouped by two-part model (hurdle vs. mixture) and approach (PICAR-Z, PICAR-Z with cross-correlation, and low-rank approach using bisquare basis functions). Results include the root mean squared prediction error (rmspe) for the entire validation dataset and the non-zero data.  For zero- vs. non-zero classification, we report the area under the ROC curve (AUC). Model-fitting walltimes are reported in minutes.} \label{Table:MacomaResults}
\end{table}

\subsection{Ice-sheet Thickness Data for West Antarctica}\label{Subsec:Antarctic}

Similar to the previous application, we partition our dataset accordingly by assigning $23,000$ locations as a training set and the remaining $6,241$ locations as the validation set. We model the observed ice thickness data using: (1) a hurdle semicontinuous model with a lognormal prevalence process; and (2) a semi-continuous mixture model using the Tobit Type I model. We rescale the spatial domain into the unit square, and include the X- and Y- coordinates as covariates. We implement similar settings as the semi-continuous cases in Section \ref{Sec:SimulationStudy} including the parameters' prior distributions. Due to the high-dimensional observations, we omit comparisons to the correlated PICAR-Z and `reparameterized' approaches. For PICAR-Z, we construct a triangular mesh with $m=5,888$ mesh vertices. For both the hurdle model, the automated heuristic chose ranks $p_{o}=46$ and $p_{p}=200$ for the occurrence and prevalence processes, respectively. In the mixture model, we use ranks $p_{o}=46$ and $p_{p}=80$. For the low-rank approach, we use the quad-tree structure (84 basis functions) from Section \ref{Sec:SimulationStudy}. 

\begin{figure}[ht]
    \centering
    \includegraphics[width=1\textwidth]{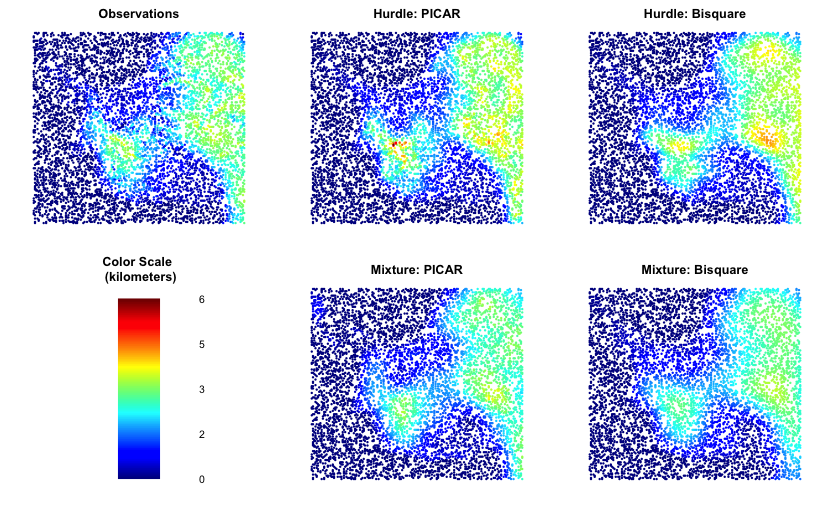}
    \caption{Maps of the observed ice-thickness at the 6,241 validation locations (top left). Predicted values using the hurdle semi-continuous model with PICAR-Z (top middle) and low-rank (bisquare) (top right). Predictions using the mixture semi-continuous model with PICAR-Z (bottom middle) and low-rank (bisquare) (bottom right).}\label{Fig:IceSheet}
\end{figure}

Comparative results are provided in the Table ~\ref{Tab:IceSheetResults} and maps of the predicted values are provided in Figure \ref{Fig:IceSheet}. Both PICAR-Z performs yields more accurate out-of-sample predictions than the low-rank approach in both the total rmspe (hurdle: $353.31$ vs. $372.53$ and mixture: $314.47$ vs. $323.77$) and the rmspe among positive values (hurdle: $436.39$ vs. $461.27$ and mixture: $384.32$ vs. $398.51$). Both approaches yield comparable results among zero vs. non-zero predictions based on AUC (hurdle: $0.95$ vs. $0.96$ and mixture: $0.94$ vs. $0.92$ ). For the hurdle model, walltimes are slightly shorter for the low-rank (bisquare) approach, which can be attributed to fewer basis functions chosen to represent the prevalence process. Despite the longer walltimes, PICAR-Z provides more accurate predictions and a larger effective sample size. The mixture model (Tobit Type I) is able to predict more accurately than the hurdle model; however, it comes at the cost of slightly longer walltimes.  Maps of the corresponding prediction standard deviations are provided in the supplement. Note that PICAR-Z yields more accurate predictions, albeit with larger uncertainties.
 
\begin{table}[ht]
\centering
\begin{tabular}{|c|c|cccc|}
  \hline
  Two-Part& Fitting & Total & Non-Zero & Zero  & Time \\ 
Model& Method & RMSPE & RMSPE & AUC & (mins) \\ 
  \hline
Semi-Continuous & PICAR-Z & 352.31 & 436.39 & 0.95 & 31.47 \\ 
Hurdle & Bisquare & 372.53 & 461.27 & 0.96 & 28.73 \\ 
  \hline
Semi-Continuous & PICAR-Z &  314.47 & 384.32 & 0.94 & 33.44 \\ 
Mixture& Bisquare & 323.77 & 398.51 & 0.92 & 34.29 \\ 
   \hline
\end{tabular}
\caption{West Antarctica Ice Thickness Results: Results are grouped by two-part model (hurdle vs. mixture) and approach (PICAR-Z vs. low-rank with bisquare basis functions). Results include the root mean squared prediction error (rmspe) for the entire validation dataset and the non-zero data.  For zero- vs. non-zero classification, we report the area under the ROC curve (AUC). Model-fitting walltimes are reported in minutes.}\label{Tab:IceSheetResults}
\end{table}

\section{Discussion}\label{Sec:Discussion}
In this study, we propose a computationally efficient approach (PICAR-Z) to model high-dimensional zero-inflated spatial count and semi-continuous observations. Our method approximates the two latent spatial random fields using the PICAR representation. PICAR-Z scales well to higher dimensions, is automated, and extends to a wide range of spatial two-part models. In both simulated and real-world examples, PICAR-Z yields comparable results to the `reparameterized' approach in both inference and prediction, yet incurs just a fraction of the computational costs. In addition, PICAR-Z outperforms a competing approach in both predictions and computational costs. Our method can be easily implemented in a programming language for Markov chain Monte Carlo algorithms such as {\tt nimble} and {\tt stan}. PICAR-Z significantly reduces computational overhead while maintaining model performance; thereby allowing practitioners to investigate a wider range of two-part spatial models than was previously possible.

While this study focuses on four types of two-part models, a natural extension would extend ideas from complex two-part models to the spatial setting. Examples include hurdle models with skewed distributions \citep{Dreassi2014Small,liu2016analyzing}, t-distributions to model heavy tailed behavior \citep{Neelon2015Semicontinuous}, or scale mixtures of normal distributions \citep{Fruhwirth2010}. Next, our approach does not provide a procedure for choosing between hurdle and mixture models, at least prior to model-fitting. Developing a formal test or automated heuristic to select the appropriate class of spatial two-part models would be a promising area of future research. 

Extending PICAR-Z to the multivariate or spatio-temporal setting has potential for future research. The latent spatial processes can be linked using nonstationary multivariate covariance functions \citep{kleiber2012nonstationary} and or basis functions weighted with Gaussian graphical vectors \citep{krock2021modeling}. For spatio-temporal data, modeling basis coefficients as a vector-autoregressive process \citep{bradley2015spatio} can induce temporal dependencies in the latent spatial occurrence and prevalence processes. In light of recent critiques from \citet{zimmerman2022deconfounding} and \citet{khan2022restricted}, subsequent studies would benefit from rigorously examining spatial confounding in zero-inflated spatial datasets by modifying PICAR-Z and the de-confounded Moran's basis functions \citep{hughes2013dimension}.

\section{Acknowledgments}\label{Sec:Acknowledgements}
We are grateful to Yawen Guan and Jaewoo Park for helpful discussions. This research received no specific grant from any funding agency in the public, commercial, or not-for-profit sectors. Any opinions, findings, and conclusions or recommendations expressed in this material are those of the authors and do not necessarily reflect the views of any funding entity. Any errors and opinions are, of course, those of the authors. We are not aware of any real or perceived conflicts of interest for any authors. This project was supported by resources provided by the Office of Research Computing at George Mason University (URL: \href{https://orc.gmu.edu}{https://orc.gmu.edu}) and funded in part by grants from the National Science Foundation (Award Number 2018631). Accompanying code can be accessed at the following github repository \url{https://github.com/benee55/PICAR_Z_Code}.

\section{Declarations}
\paragraph{Funding and/or Conflicts of interests/Competing interests}
This research received no specific grant from any funding agency in the public, commercial, or not-for-profit sectors. All authors declare that they have no conflicts of interest.

\newpage
\bibliography{references}

\end{document}


\doublespacing
\maketitle

\section{Two-Part Models for Zero-inflated Data}
In this section, we review the two-part modeling framework \citep{Mullahy1986,lambert1992zero} for zero-inflated data. Two-part models are comprised of two random variables: (1) the occurrence random variable $O$ that generates the structural zero and non-zeros cases; and (2) the prevalence random variable $P$, which assigns positive values for the structural non-zero cases. In special cases, the prevalence random variable $P$ can also generate zeros. The zero-inflated observation $Z$ is generated as follows:
\begin{equation}\label{EQ:TwoPartGenerating}
    Z =
\left\{
	\begin{array}{ll}
		0  & \mbox{if } O = 0 \\
		P & \mbox{if } O = 1.
	\end{array},
\right.
\end{equation}
where $O$ and $P$ are the latent occurrence and prevalence random variables, respectively. The occurrence variable $O\in\{0,1\}$ is typically modeled as a Bernoulli random variable with probability $\pi\in(0,1)$ (i.e., $O \sim \mbox{Bern}(\pi)$). The prevalence variable is distributed as $P\sim F(\btheta)$ where $F(\btheta)$ is a discrete probability mass function or continuous probability density function with prevalence model parameter $\btheta$. 

Two-part models for zero-inflated data typically fall into two classes - hurdle and mixture models.  In \textit{hurdle models}, the occurrence random variable specifies which data have zero- or non-zero values. For the non-zero data, their respective positive values are generated via the prevalence random variable $P$. In the discrete case, $F(\cdot)$ is a zero-truncated probability mass function such as the zero-truncated Poisson or the zero-truncated negative binomial distribution. For semi-continuous observations, $F(\cdot)$ is a probability density function with positive support such as a log-normal or gamma distribution. For \textit{mixture models}, the zero-valued observations can be generated by both the occurrence $O$ and prevalence random variables $P$. Here, $O$ determines whether a data point is classified as a structural zero or structural non-zero case. For the structural non-zero cases, the prevalence random variable $P$ can generates both zeros and positive values. In the discrete case, $F(\cdot)$ is a non-degenerate mass function such as the Poisson or Negative-Binomial distribution. For semi-continuous observations, $F(\cdot)$ can be a censored model such as a Tobit Type I. 

The probability mass (or density) function $f_{Z}(z)$ for two-part models is as follows:
\begin{equation}
    f_{Z}(z) =
\left\{
	\begin{array}{ll}
		(1-\pi) +\pi \tilde{f}(0),  & \mbox{if } z=0 \\
		\pi \tilde{f}(z), & \mbox{if } z>0.
	\end{array},
\right.
\end{equation}
where $\pi\in(0,1)$ is occurrence probability, $\tilde{f}(z)$ is a probability mass or density function for the prevalence random variable with parameters $\btheta$. For two part models, the expectation is defined as $E[Z|\pi,\btheta] = \pi E_{\tilde{f}}[Z|\btheta]$ and the variance is $Var[Z|\pi,\btheta]= \pi(1- \pi)E_{\tilde{f}}[Z|\btheta]^{2} + \pi Var_{\tilde{f}}[Z|\btheta]$ where $E_{\tilde{f}}[Z|\btheta]$ and $Var_{\tilde{f}}[Z|\btheta]$ denote the expectation and variance of a random variable with probability mass or density function $\tilde{f}(\cdot)$ and parameter $\btheta$. 

Alternative distributions of $\tilde{F}(\cdot)$ can result in richer and more flexible two-part models. For count data, examples include the Poisson, negative binomial, zero-truncated Poisson \citep{lambert1992zero}, translated Poisson \citep{Hoef2007Seals}, zero-truncated negative binomial \citep{mwalili2008zero}, generalized Poisson \citep{Czado2008}, and binomial distributions \citep{hall2000zero}. In the semi-continuous case, the lognormal distribution may not be appropriate due to the lack of symmetry or fatter tails exhibited by the observations. Past studies have used skewed distributions \citep{Dreassi2014Small,liu2016analyzing}, t-distributions to model heavy tailed behavior \citep{Neelon2015Semicontinuous}, or modeled the prevalence process using scale mixtures of normal distributions \citep{Fruhwirth2010}.

Since two-part models have a natural hierarchical structure (Equation \ref{EQ:TwoPartGenerating}), they can be readily incorporated into the Bayesian hierarchical modeling framework. The general Bayesian hierarchical model is as follows:
\begin{align*}
\textbf{Data Model: } & \qquad Z|O,P \sim F\big(\cdot|O,P\big)\\
 \textbf{Process  Model: } & \qquad O|\pi \sim \mbox{Bern}(\pi)\\
 & \qquad  P|\theta \sim \tilde{F}(\theta)\\
\textbf{Parameter  Model: }& \qquad \mbox{Priors for } \pi \mbox{ and } \theta
\end{align*}

\subsection{Spatial Generalized Linear Mixed Models}\label{SubSec:SGLMSS}

In the literature, the spatially-dependent occurrence $O(\bs)$ and prevalence $P(\bs)$ processes have been modeled as spatial generalized linear mixed models (SGLMMs)\citep[cf.][]{Agarwal2002Zero, Rathbun2006Spatial, Neelon2013Poisson, recta2012two}. Spatial generalized linear mixed models are a popular choice for modeling spatial data \citep{diggle1998model, haran2011gaussian}, particularly non-Gaussian data. Let $\{Z(\bs):\bs\in \mathcal{D}\}$ be a non-Gaussian spatial random process. Assuming the realizations $Z(\bs)$ are conditionally independent given the latent spatial random effects $W(\bs)$, the conditional mean $E[Z(\bs)|\pmb{\beta},W(\bs),\epsilon(\bs)]$ can be modeled through the linear predictor $\eta(\bs)$:
$$\eta(\bs)=g\{E[Z(\bs)|\bbe,W(\bs),\epsilon(\bs)]\}=\bX(\bs)^{\prime}\bbe +W(\bs)+\epsilon(\bs),$$
where $g(\cdot)$ is a known link function and $\epsilon(\bs)$ is micro-scale measurement error typically modeled as an uncorrelated Gaussian process where $\epsilon(\bs)\sim N(0,\tau^2)$ for $\bs \in \mathcal{D}$. Binary and count observations are two common types of non-Gaussian spatial data, and these can be modeled using the binary SGLMM with logit link and the Poisson SGLMM with log link, respectively. 
The general Bayesian hierarchical framework for the SGLMM is:
\begin{align*}
    \mbox{Data Model: }&\qquad Z(\bs)|\mu(\bs) \sim F\big(\cdot|\mu(\bs)\big), \qquad \mu(\bs)=\mathbb{E}[Z(\bs)|\pmb{\beta},W(\bs),\epsilon(\bs)]\\
    & \qquad \eta(\bs) = g(\mu(\bs))= \bX(\bs)^{\prime}\pmb{\beta} + W(\bs) + \epsilon(\bs)\\
    \mbox{Process  Model: } & \qquad \bW|\bTheta \sim N(\bzero,\bC(\bTheta)), \qquad \mathbf{W}=(W(\bs_{1}),..,W(\bs_{n}))'\\
    & \qquad \epsilon(\bs)|\tau^{2} \sim \mathcal{N}(0, \tau^{2})\\
        \mbox{Parameter  Model: } & \qquad  \pmb{\beta}\sim p(\pmb{\beta}),\quad \pmb{\Theta}\sim p(\pmb{\Theta}),\quad \tau^{2}\sim p(\tau^{2})
\end{align*}
where  $F\big(\cdot|\mu(\bs)\big)$ is a probability distribution (e.g. Bernoulli or Poisson), $\bbe$ and $\bX(\bs)$ are the $k$-dimensional vector of the fixed effects and covariates for location $\bs_{i}$, respectively. $\mathbf{W}$ is the $n$-dimensional vector of the spatial random effects, and $\tau^{2}$ is the nugget variance. $\bC(\bTheta)$ is the covariance matrix for the spatial random effects $\bW$ with covariance parameter $\bTheta$.

To account for spatial dependence, $\mathbf{W}=\{W(\bs): \bs\in \mathcal{D}\}$ can be modeled as a stationary zero-mean Gaussian process with a positive definite covariance function $C(\cdot)$. For a finite set of locations, the spatial random effects $\mathbf{W}$ follow a multivariate normal distribution $\mathbf{W}|\Theta \sim N(\bzero,\bC(\bTheta))$ with covariance matrix $\bC(\bTheta)$ defined such that $\bC(\bTheta)_{ij}=\mbox{Cov}(W(\bs_{i}),W(\bs_{j}))$ and covariance function parameter $\bTheta$. The Mat\'ern covariance function is a widely used class of stationary and isotropic covariance functions \citep{stein2012interpolation} defined as:
\begin{equation}\label{Eq:Matern}
  \bC(\bs_{i},\bs_{j})=\sigma^{2}\frac{1}{\Gamma(\nu)2^{\nu-1}}\Big(\sqrt(2\nu)\frac{h}{\phi}\Big)^{\nu}K_{\nu}\Big(\sqrt(2\nu)\frac{h}{\phi}\Big)  
\end{equation}
where $h=||\bs_{i}-\bs_{j}||$ is the Euclidean distance between locations $\bs_{i}$ and $\bs_{j}$, $\sigma^{2}>0$ is the partial sill or scale parameter of the process, and $\phi>0$ is the range parameter for spatial dependence. $K_{\nu}(\cdot)$ is the modified Bessel function of the second kind where the smoothness parameter $\nu$ is commonly fixed prior to model fitting.

\section{Types of Spatial Two-part Models}
Here, we introduce four commonly used spatial two-part models and provide a summary in Table \ref{Table:TwoPart}.

\subsubsection*{Hurdle Model for Spatial Count Data}
The hurdle count model is a sensible two-part model for spatial zero-inflated data. This model assumes the occurrence process $O(\bs)$ solely generates zero-valued data; hence, the distribution of the prevalence process $\tilde{F}(\cdot|\theta(\bs))$ is typically zero-truncated distribution such as a zero-truncated Poisson distribution, zero-truncated Negative binomial, or a translated Poisson distribution \citep{Hoef2007Seals}. For the zero-truncated Poisson distribution, the likelihood function of the hurdle count model is defined as:
\begin{equation}\label{EQ:H_Count}
    f\big(z|O(\bs),P(\bs)\big) =
\left\{
	\begin{array}{ll}
		\pi(\bs),  & \mbox{if } z=0 \\
		(1-\pi(\bs))\times \frac{\theta(\bs)^{z}e^{-\theta(\bs)}}{z!(1-e^{-\theta(\bs)})}, & \mbox{if } z>0.
	\end{array}.
\right.
\end{equation}
where $\pi(\bs)= g_o^{-1}(\bX(\bs)^{\prime}\bbe_{o} + W_{o}(\bs) +\epsilon_{o}(\bs))$ is the occurrence probability process, and $\theta(\bs)=\exp\{\bX(\bs)^{\prime}\bbe_P+\bW_p(\bs)+\epsilon_{p}(\bs)\}$ is the intensity process. 

\subsubsection*{Hurdle Model for Spatial Semi-continuous Data}
For semi-continuous observations, an appropriate two-part model is a hurdle model based on a continuous distribution over strictly positive real numbers. Similar to the discrete case, the hurdle model assumes the zero-valued data are generated from occurrence process $O(\bs)$ only. Common choices for the distribution of $P(\bs)$ are the lognormal, skewed-normal, truncated normal, and log-skew-t \citep{neelon2016modeling, chai2008use}. A popular example is the Bernoulli-lognormal hurdle model with likelihood function:
\begin{equation}\label{EQ:H_Semi}
    f\big(z|O(\bs),P(\bs)\big) =
\left\{
	\begin{array}{ll}
		\pi(\bs),  & \mbox{if } z=0 \\
		(1-\pi(\bs))\times LN(z|\theta(\bs)), & \mbox{if } z>0,
	\end{array}.
\right.
\end{equation}
where $LN(z|\theta(\bs))= \frac {1}{z {\sqrt {2\pi\tau^{2} }}}\ \exp \left(-{\frac {\left(\ln z-\mu(\bs) \right)^{2}}{2\tau^{2}}}\right)$ is the probability density function of a lognormal distribution with parameters $\theta(\bs)=\{\mu(\bs),\tau^{2}\}$. $\pi(\bs)$ can be modeled similarly as in the hurdle count model. However, the spatially-varying mean process is modeled as $\mu(\bs)=\exp(\bX(\bs)^{\prime}\bbe_P+\bW_p(\bs)+\epsilon_{p}(\bs))$.

\subsubsection*{Mixture Model for Spatial Count Data}
The spatial zero-inflated mixture model is appropriate for modeling zero-inflated count data where both the occurrence $O(\bs)$ and prevalence processes $P(\bs)$ generate zero-values. $\tilde{F}(\cdot|\theta(\bs))$ is typically a discrete distribution whose support is the non-negative integers (e.g. Poisson or Negative-Binomial distribution). A commonly-used case is the zero-inflated Poisson (ZIP) model, which utilizes the Poisson distribution. The corresponding likelihood function is:
\begin{equation}\label{EQ:Mix_Count}
    f\big(z|O(\bs),P(\bs)\big) =
\left\{
	\begin{array}{ll}
		\pi(\bs) +(1-\pi(\bs)) e^{-\theta(\bs)},  & \mbox{if } z=0 \\
		(1-\pi(\bs)) \frac{\theta(\bs)^{z}e^{-\theta(\bs)}}{z!}, & \mbox{if } z>0.
	\end{array}.
\right.
\end{equation}
where $\pi(\bs)$ is modeled similarly as in the previous cases, and the spatially-varying intensity process $\theta(\bs)$ is modeled as $\theta(\bs)=\exp\{\bX(\bs)^{\prime}\bbe_P+\bW_p(\bs)+\epsilon_{p}(\bs)\}$. 

\subsubsection*{Mixture Model for Spatial Semi-continuous Data}
Mixture models can also be extended to spatial zero-inflated semi-continuous data where both the occurrence $O(\bs)$ and prevalence processes $P(\bs)$ generate zeros. For this class of models, the prevalence process $P(\bs)$ generally includes a censored distribution such as a Type 1 Tobit model \citep{tobin1958estimation,moulton1995mixture} to generate `censored' zeros and positive values. One example is the zero-inflated Tobit (ZIT) two-part model. The Type 1 Tobit model generates censored data  $P(\bs)\in\{0,\mathbb{R}^{+}\}$ as follows:

\begin{equation}
    P(\bs) =
\left\{
	\begin{array}{ll}
	    P^{*}(\bs),  & \mbox{if } P^{*}(\bs)>\gamma \\
		0, & \mbox{if } P^{*}(\bs)\leq \gamma
	\end{array},
\right.
\end{equation}
where $\gamma$ is a threshold (or lower detection limit) and $P^{*}(\bs)$ is an auxiliary latent random variable where $P^{*}(\bs)\sim \mathcal{N}(\mu(\bs),\tau^{2})$. For a threshold $\gamma$, we have the following likelihood function $f_{Z}\big(z;\theta(s)\big)$:
\begin{equation}\label{EQ:Tobit}
    f_{Z}\big(z;\theta\big) =
\left\{
	\begin{array}{ll}
	    \Phi(\frac{\gamma-\mu(s)}{\tau}),  & \mbox{if } z=0 \\
		\frac{1}{\tau}\phi(\frac{z-\mu(s)}{\tau}), & \mbox{if } z>0.
	\end{array},
\right.
\end{equation}
where $\Phi(\cdot)$ is the standard normal cumulative distribution function and $\phi(\cdot)$ is the standard normal probability density function. The estimable model parameters are $\theta(s)=\{\mu(s),\tau^{2}\}$, and the likelihood function for the zero-inflated Tobit (ZIT) model is:
\begin{equation}\label{EQ:Mix_Semi}
    f\big(z|O(\bs),P(\bs)\big) =
\left\{
	\begin{array}{ll}
		\pi(\bs) +(1-\pi(\bs))\times \Phi(\frac{\gamma-\mu(\bs)}{\tau}),  & \mbox{if } z=0 \\
		(1-\pi(\bs))\times \frac{1}{\tau}\phi(\frac{z-\mu(\bs)}{\tau}), & \mbox{if } z>0.
	\end{array},
\right.
\end{equation}
where $\Phi(\cdot)$ and $\phi(\cdot)$ are the cumulative distribution function and the probability density function of a standard normal distribution. We model the spatially-varying occurrence process $\pi(\bs)$ in the same way as in the previous cases, and the mean process is modeled as $\mu(\bs)=\bX(\bs)^{\prime}\bbe_P+\bW_p(\bs)+\epsilon_{p}(\bs)$.

\begin{table}
\caption{Summary of Spatial Two-part Models}
\centering
\begin{tabular}{ |c|c|c|c| } 
\hline
Class & Data Type & Occurrence $O(s)$ & Prevalence $P(s)$\\
 \hline
  \multirow{4}{4.5em}{Hurdle} & \multirow{2}{5em}{Discrete} & Bernouilli & Zero-Truncated Poisson \\ 
 &  & Bernouilli & Zero-Truncated Neg. Binomial \\ 
 \cline{2-4}
 & \multirow{2}{5em}{Continuous} & Bernouilli & Lognormal \\ 
 &  & Bernouilli & Log skew-normal\\ 
  \hline
    \multirow{3}{4.5em}{Mixture} & \multirow{2}{5em}{Discrete} & Bernouilli & Poisson \\
 &  & Bernouilli & Negative Binomial \\ 
 \cline{2-4}
 & Continuous & Bernouilli & Tobit Type I Model \\ 
 \hline
\end{tabular}
\caption{Summary of Common Spatial Two-part models. Models are broken down by class (hurdle vs. mixture) and data type (discrete vs. semi-continuous).}\label{Table:TwoPart}
\end{table}

\section{Competing Approaches}
In both the simulation studies, we compare the PICAR-Z approach to two competing methods, low-rank (bisquare) approach \citep{cressie2008fixed, sengupta2013hierarchical} and a `reparameterized' method \citep{christensen2002bayesian}. We provide a brief overview and implementation details specific to our study. 

\subsection{Low-rank (bisquare) Approach}
We employ the bisquare basis function with a quad-tree structure from \citet{sengupta2013hierarchical}. We use a hierarchical structure of bisquare basis functions centered at fixed knot locations. Let $\pmb{\Phi}_{\bc_{i}}(\bs)$ denote the bisquare basis function centered at knot location $\bc_{i}$ with support on $\bs\in \mathcal{D}$, the spatial domain. The general form of the bisquare basis function is:
$$\pmb{\Phi}_{\bc_{i}}(\bs)=\Bigg\{1-\Bigg(\frac{||\bs-\bc_{i}||}{\omega}\Bigg)^{2}\Bigg\}^{2} I(||\bs-\bc_{i}||<\omega),$$
where $\bs$ is an observation location and $\bc_{i}$ is the $i$-th knot location (i.e. center of the basis function). The bi-square basis functions have locally bounded support $\{\bs \in \mathbb{R}^{2}: ||\bs-\bc_{i}||<\omega\}$ that is controlled by the `aperture', or threshold, $\omega>0$.

As in past studies \citep[cf.][]{sengupta2013hierarchical, sengupta2016predictive, shi2017spatial}, we employ the quad-tree structure to select the centers (knots) of the bisquare basis functions; thereby ensuring that the knots for different resolutions do not overlap. Moreover, we placed knots outside the observed spatial domain \cite{cressie2010high} to account for boundary effects. In this study, we use three resolutions $J=3$, where there are four knot locations $\bc_{1},...,\bc_{4}$ for the first resolution, 16 knots $\bc_{5},...,\bc_{20}$ for the second resolution, and 64 knots $\bc_{21},...,\bc_{84}$ for the third resolution. The resolution-specific ``apertures'' (thresholds) $\omega_{j}$ for $j=1,...,J$ are defined to be $1.5$ times the shortest distance between like-resolution knot locations. We constructed a set of $84$ bisquare basis functions $\pmb{\Phi}_{\bc_{i}}(\bs)$ for $\bc_{i}\in{\bc_{1},...,\bc_{84}}$ allocated proportionally across the three resolutions.

We can embed the multi-resolution bisquare basis functions into the Bayesian hierarchical framework for spatial two-part models as follows:

\begin{align*}
\textbf{Data Model: } & \qquad Z(\bs)|O(\bs),P(\bs) \sim F\big(\cdot|O(\bs),P(\bs)\big)\\
\textbf{Process Model: } & \qquad O(\bs)|\pi(\bs) \sim \mbox{Bern}(\pi(\bs))\\
& \qquad P(\bs)|\btheta(\bs) \sim \tilde{F}(\cdot | \btheta(\bs))\\
\textbf{Sub-process  Model 1: }& \qquad \pi(\bs)|\eta_{o}(\bs)=g^{-1}_{o}(\eta_{o}(\bs))\\
\textbf{(Occurrence) } & \qquad \eta_{o}(\bs)|\bbe_{o},\bdel_{o}(\bs)= \bX(s)^{\prime}\bbe_{o} + \bPhi(s)^{\prime}\bdel_{o}(\bs)\\
& \qquad \bdel_{o}|\tau^2_{o} \sim \mathcal{N}(\bzero,\tau_{o}^{2}\bSig_{o})\\
\textbf{Sub-process  Model 2: } & \qquad \theta(\bs)|\eta_{p}(\bs)=g^{-1}_{p}(\eta_{p}(\bs))\\
\textbf{(Prevalence)} & \qquad  \eta_{p}(\bs)|\bbe_{p},\bdel_{p}(\bs)=\bX(s)^{\prime}\bbe_{p} + \bPhi(s)^{\prime}\bdel_{p}(\bs) \\
& \qquad \bdel_{p}|\tau^2_{p} \sim \mathcal{N}(\bzero,\tau_{p}^{2}\bSig_{p})\\
\textbf{Parameter  Model: } & \qquad \bbe_{o}\sim \mcN(\bmu_{\beta_o}, \Sigma_{\beta_o}),\quad \bbe_{p}\sim \mcN(\bmu_{\beta_p}, \Sigma_{\beta_p})\\
& \qquad \tau_o^2 \sim \mathcal{IG}(\alpha_{\tau_o},\beta_{\tau_o}), \quad \tau_p^2 \sim \mathcal{IG}(\alpha_{\tau_p},\beta_{\tau_p})
\end{align*}

where $\bPhi$ denotes the matrix of the multi-resolution bi-square basis functions defined as:
$$
\pmb{\Phi}=\Bigg[\pmb{\Phi}_{c_{1}}\quad \pmb{\Phi}_{c_{2}}\quad  \cdots \quad  \pmb{\Phi}_{c_{84}}\Bigg]
$$
In this study, we implement the three-resolution structure with $84$ total basis functions. The hierarchical model is completed by assigning prior distributions for the model parameters $\bbe_{o}$ , $\bbe_{p}$, $\sigma^{2}_{o}$  and $\sigma^{2}_{p}$. In the simulation study, we use the following prior distributions for the model parameters, $\bbe_{o}\sim \mcN(\bzero, 100\bI)$, $\bbe_{p}\sim \mcN(\bzero, 100\bI)$, $\tau_o^2\sim \mathcal{IG}(0.002, 0.002)$, and $\tau_p^2\sim \mathcal{IG}(0.002, 0.002)$

\subsection{Reparameterization}
Reparameterization approaches \citep{christensen2002bayesian, haran2003accelerating, Guan_Haran_2018} have been designed to reduce correlation in the spatial random effects, which often results in faster mixing MCMC algorithms. However, these techniques can be very expensive for high-dimensional data, since the reparameterization step is still requires an expensive Cholesky decomposition $\mathcal{O}(n^3)$ and the number of spatial random effects remain unchanged.

We employ a the reparameterization approach from \citet{christensen2002bayesian} where the latent spatial processes are reparameterized as $\mathbf{W}_{o}=\mathbf{L}_{o}\bga_{o}$ and $\mathbf{W}_{p}=\mathbf{L}_{p}\bga_{p}$ where $\mathbf{L}_{o}$ and $\mathbf{L}_{p}$ are the lower triangular matrices obtained through the Cholesky decomposition of the corresponding covariance matrices (i.e., $C_{o}(h;\sigma_{o}^{2},\phi_{o})=\mathbf{L}_{o}\mathbf{L}_{o}'$ and $C_{p}(h;\sigma_{p}^{2},\phi_{p})=\mathbf{L}_{p}\mathbf{L}_{p}'$). For strongly correlated spatial random effects, this approach results in approximately independent reparameterized spatial random effects $\gamma_{o}$ and $\gamma_{p}$ for the occurrence and prevalence processes, respectively; thereby improving mixing in MCMC algorithms \citep{christensen2002bayesian, Christensen2006robust}. 
\newpage
Under the reparameterization framework, the Bayesian hierarchical model for spatial two-part models is:
\begin{align}\label{EQ:ReparameterizedTwoPart}
    \textbf{Data Model:}&\qquad \bZ|O(\bs),P(\bs) \sim F\big(\cdot|O(\bs),P(\bs)\big) \\
    \textbf{Process  Model:} & \qquad O(\bs)|\pi(\bs) \sim \mbox{Bern}(\cdot|\pi(\bs)) \nonumber \\
    & \qquad P(\bs)|\btheta(\bs) \sim \tilde{F}(\cdot|\btheta(\bs))\nonumber \\
\textbf{Sub-process  Model 1: } & \qquad \pi(\bs)|\eta_{o}(\bs)=g^{-1}_{o}(\eta_{o}(\bs))\nonumber\\
\textbf{(Occurrence) } & \qquad \eta_{o}(\bs)|\bbe_{o},\bga_{o} =\bX(\bs)^{\prime}\bbe_{o} + [\bL_{\phi_{o}} \bga_{o}](\bs)  \nonumber\\
& \qquad \bR_{\phi_{o}}=\bL_{\phi_{o}}\bL_{\phi_{o}}^{\prime}\nonumber \\
& \qquad \bga_{o}|\sigma^{2}_{o}\sim \mathcal{N}(\pmb{0},\sigma^{2}_{o}\bI), \nonumber\\
\textbf{Sub-process  Model 2: } & \qquad \theta(\bs)|\eta_{p}(\bs)=g^{-1}_{p}(\eta_{p}(\bs)) \nonumber\\
\textbf{(Prevalence)}  & \qquad \eta_{p}(\bs)|\bbe_{p},\bga_{p}=\bX(\bs)^{\prime}\bbe_{p} + [\bL_{\phi_{p}} \bga_{p}](\bs) \nonumber\\
& \qquad \bR_{\phi_{p}}=\bL_{\phi_{p}}\bL_{\phi_{p}}^{\prime}\nonumber \\
& \qquad \bga_{p}|\sigma^{2}_{p}\sim \mathcal{N}(\pmb{0},\sigma^{2}_{p}\bI), \nonumber\\
\textbf{Parameter  Model:} &\qquad  \bbe_{o}\sim p(\bbe_{o}),\quad \bbe_{p}\sim p(\bbe_{p}), \quad \phi_{o} \sim p(\phi_{o}) , \quad \phi_{p} \sim p(\phi_{p}) \nonumber\\
& \qquad \sigma^{2}_{o} \sim p(\sigma^{2}_{o}), \quad \sigma^{2}_{p} \sim p(\sigma^{2}_{p})
\end{align}
where $F\big(\cdot|O(\bs),P(\bs)\big)$ is the distribution function of a spatial two-part model. $\bR_{\phi_{p}}$ and $\bR_{\phi_{p}}$ are the correlation matrices for the spatial occurrence and prevalence processes, respectively. For this approach, we assume that the class of the Mat\'ern correlation function is known ($\nu=0.5$). $\phi_{o}$ and $\phi_{p}$ denote the range parameters for the occurrence and prevalence cases. A Cholesky decomposition of $\bR_{\phi_{p}}$ and $\bR_{\phi_{p}}$ result in lower triangular matrices $\bL_{\phi_{o}}$ and $\bL_{\phi_{p}}$. 

In the simulation study, we use the following prior distributions for the model parameters, $\bbe_{o}\sim \mcN(\bzero, 100\bI)$, $\bbe_{p}\sim \mcN(\bzero, 100\bI)$, $\phi_{o}\sim \mathcal{U}(0,\sqrt{2})$, $\phi_{p}\sim \mathcal{U}(0,\sqrt{2})$, $\ssq_{o}\sim \mathcal{IG}(0.002, 0.002)$, and $\ssq_{p}\sim \mathcal{IG}(0.002, 0.002)$

\begin{table}[ht]
\centering
\begin{tabular}{rrrrr}
  \hline
   & & Correlated& Low-Rank &   \\ 
 & PICAR-Z &  PICAR-Z & (Bisquare) &  Reparameterized \\ 
  \hline
Count Hurdle & 100.0 & 74.9 & 49.4 & 0.7 \\ 
  Semi-continuous Hurdle & 79.3 & 52.9 & 62.8 & 0.7 \\ 
  Count Mixture & 49.2 & 40.1 & 17.0 & 0.3 \\ 
  Semi-continuous Mixture & 24.1 & 19.1 & 10.9 & 0.1 \\ 
   \hline
\end{tabular}
\caption{Average effective samples per second (ESS/sec) for the coefficients $\bbe_0$ and $\bbe_P$ from the simulation study. The rows pertain to the spatial two-part model used to generate the simulated observations. The columns denote the four methods implemented in the simulation study - PICAR-Z, correlated PICAR-Z, low-rank (bisquare), and the `reparameterized' approach. The ESS/sec values are averaged over all 100 samples for the corresponding two-part model.}
\end{table}

\begin{table}[ht]
\centering
\begin{tabular}{rrrrr}
  \hline
   & & Correlated& Low-Rank &   \\ 
 & PICAR-Z &  PICAR-Z & (Bisquare) &  Reparameterized \\ 
  \hline
$\beta_{1O}$ & 0.94 & 0.94 & 0.83 & 0.94 \\ 
$\beta_{2O}$ & 0.96 & 0.95 & 0.88 & 0.97 \\ 
$\beta_{1P}$  & 0.75 & 0.74 & 0.86 & 0.86 \\ 
$\beta_{2P}$ & 0.82 & 0.83 & 0.89 & 0.90 \\ 
   \hline
\end{tabular}
\caption{Count Hurdle Case: Coverage probabilities for the coefficients $\bbe_0$ and $\bbe_P$ from the simulation study. The rows pertain to the four regression coefficients - $\beta_{1O}$, $\beta_{2O}$, $\beta_{1P}$, and $\beta_{2P}$. The columns denote the four methods implemented in the simulation study - PICAR-Z, correlated PICAR-Z, low-rank (bisquare), and the `reparameterized' approach. Coverage probabilities are computed using the 100 simulated samples.}
\end{table}

\begin{table}[ht]
\centering
\begin{tabular}{rrrrr}
  \hline
   & & Correlated& Low-Rank &   \\ 
 & PICAR-Z &  PICAR-Z & (Bisquare) &  Reparameterized \\ 
  \hline
$\beta_{1O}$ & 0.92 & 0.93 & 0.82 & 0.94 \\ 
$\beta_{2O}$ & 0.92 & 0.94 & 0.87 & 0.93 \\ 
$\beta_{1P}$ & 0.88 & 0.87 & 0.91 & 0.75 \\ 
$\beta_{2P}$ & 0.90 & 0.92 & 0.91 & 0.85 \\ 
\hline
\end{tabular}
\caption{Semi-continuous Hurdle Case: Coverage probabilities for the coefficients $\bbe_0$ and $\bbe_P$ from the simulation study. The rows pertain to the four regression coefficients - $\beta_{1O}$, $\beta_{2O}$, $\beta_{1P}$, and $\beta_{2P}$. The columns denote the four methods implemented in the simulation study - PICAR-Z, correlated PICAR-Z, low-rank (bisquare), and the `reparameterized' approach. Coverage probabilities are computed using the 100 simulated samples.}
\end{table}

\begin{table}[ht]
\centering
\begin{tabular}{rrrrr}
  \hline
   & & Correlated& Low-Rank &   \\ 
 & PICAR-Z &  PICAR-Z & (Bisquare) &  Reparameterized \\ 
  \hline
$\beta_{1O}$ & 0.89 & 0.87 & 0.84 & 0.90 \\ 
$\beta_{2O}$ & 0.94 & 0.96 & 0.90 & 0.89 \\ 
$\beta_{1P}$  & 0.72 & 0.71 & 0.63 & 0.88 \\ 
$\beta_{2P}$ & 0.65 & 0.67 & 0.62 & 0.86 \\ 
\hline
\end{tabular}
\caption{Count Mixture Case: Coverage probabilities for the coefficients $\bbe_0$ and $\bbe_P$ from the simulation study. The rows pertain to the four regression coefficients - $\beta_{1O}$, $\beta_{2O}$, $\beta_{1P}$, and $\beta_{2P}$. The columns denote the four methods implemented in the simulation study - PICAR-Z, correlated PICAR-Z, low-rank (bisquare), and the `reparameterized' approach. Coverage probabilities are computed using the 100 simulated samples.}
\end{table}

\begin{table}[ht]
\centering
\begin{tabular}{rrrrr}
  \hline
   & & Correlated& Low-Rank &   \\ 
 & PICAR-Z &  PICAR-Z & (Bisquare) &  Reparameterized \\ 
  \hline
$\beta_{1O}$ & 0.11 & 0.06 & 0.40 & 0.54 \\ 
$\beta_{2O}$ & 0.12 & 0.09 & 0.32 & 0.53 \\ 
$\beta_{1P}$ & 0.53 & 0.50 & 0.33 & 0.50 \\ 
$\beta_{2P}$ & 0.52 & 0.49 & 0.43 & 0.41 \\ 
\hline
\end{tabular}
\caption{Semi-continuous Mixture Case: Coverage probabilities for the coefficients $\bbe_0$ and $\bbe_P$ from the simulation study. The rows pertain to the four regression coefficients - $\beta_{1O}$, $\beta_{2O}$, $\beta_{1P}$, and $\beta_{2P}$. The columns denote the four methods implemented in the simulation study - PICAR-Z, correlated PICAR-Z, low-rank (bisquare), and the `reparameterized' approach. Coverage probabilities are computed using the 100 simulated samples.}
\end{table}

\section{Estimation of Cross-Correlation}
In this section, we propose a modification of the PICAR-Z approach to model the cross-covariance between the occurrence $O(\bs)$ and prevalence processes $P(\bs)$. We extend the general modeling framework from \citet{recta2012two} to high-dimensional settings by imposing correlation on the dimension-reduced PICAR basis coefficients $\bdel_{o}$ and $\bdel_{p}$.

Suppose we have realizations $\bW_{o}=(W_{o}(\bs_1),...,W_{o}(\bs_1))^{\prime}$ and $\bW_{p}=(W_{p}(\bs_1),...,W_{p}(\bs_1))^{\prime}$ from the occurrence and prevalence processes, respectively. Past studies \citep{recta2012two,oliver2003gaussian} have modeled cross-covariance between two spatial processes as:

\begin{equation}\label{EQ:RandomEffectsCross}
    \begin{bmatrix}
\bW_{O}\\
\bW_{P}
\end{bmatrix} |\tau_{o},\ssq_{p},\phi_{o},\phi_{p}  \sim \mathcal{N}\Bigg(
\begin{bmatrix}
\bzero\\
\bzero
\end{bmatrix}
, 
\begin{bmatrix}
\ssq_{o}\bR_{\phi_{o}} & \rho \bSig_{op}\\
\rho \bSig_{op}^{\prime} & \ssq_{p}\bR_{\phi_{p}}\\
\end{bmatrix}
\Bigg).
\end{equation}

where $\bW_{O}$ and $\bW_{p}$ are realizations from a spatial occurrence and prevalence processes. $\ssq_{o}\bR_{\phi_{o}}$ and $\ssq_{p}\bR_{\phi_{p}}$ are the spatial covariance matrices for the spatial occurrence and prevalence processes with marginal variance parameters $\ssq_{o}$ and $\ssq_{o}$ and range parameters $\phi_{o}$ and $\phi_{p}$. In this study, we utilize the Mat\'ern class of covariance functions with smoothness $\nu=0.5$, whic results in the exponential covariance function. Here, the $(i,j)$-th element of the covariance matrix is $\ssq_{o}\bR_{\phi_{o}ij}=\ssq_{o}\exp\{-||\bs_{i}-\bs_{j}||/\phi_{o}\}$ and $\ssq_{p}\bR_{\phi_{p}ij}=\ssq_{p}\exp\{-||\bs_{i}-\bs_{j}||/\phi_{p}\}$. We assume that the covariance is isotropic and stationary; hence, the covariance is merely  a function of the Euclidean distance between the two locations. The cross-covariance function is constructed by setting $\rho\bSig_{op}=\rho\bL_{o}\bL_{p}$ as in \citet{oliver2003gaussian} where $\rho$ is the correlation between the occurrence and prevalence processes (at the same location) and $\bL_{o}$ and $\bL_{p}$ are the Choleski factors of $\ssq_{o}\bR_{\phi_{o}}$ and $\ssq_{p}\bR_{\phi_{p}}$, respectively. That is,  $\ssq_{o}\bR_{\phi_{o}}=\ssq_{o}\bL_{o}\bL_{o}^{\prime}$ and $\ssq_{p}\bR_{\phi_{p}}=\ssq_{p}\bL_{p}\bL_{p}^{\prime}$. 

Though this modeling framework has been successfully applied to \citet{recta2012two} to smaller datasets, it may be computationally prohibitive for larger datasets. For a sample of $n=400$ observations, inference required 60 hours to generate 100,000 iterations of the Langevin-Hastings algorithm on a 3.0-GHz quadcore Intel Xeon processor with 32 GB of memory \citep{recta2012two}. 

Instead, we propose modeling the cross-correlation among the PICAR-Z basis coefficients $\bdel_{o}$ and $\bdel_{p}$, which results in cross-correlation between the resulting spatial random effects $\tilde{\bW}_{o}\approx \bPhi_{o}\bdel_{o}$ and $\tilde{\bW}_{p}\approx \bPhi_{p}\bdel_{p}$ for the occurrence and prevalence processes, respectively.

By construction, the matrices $\bPhi_{P}\bPhi_{P}^{\prime}$ and $\bPhi_{O}\bPhi_{O}^{\prime}$ are positive semi-definite with Cholesky factors $L_1$ and $L_2$, respectively. If $(\bdel_O^{\prime},\bdel_P^{\prime})^{\prime}$ is distributed as:

$$\begin{bmatrix}
\bdel_O\\
\bdel_P
\end{bmatrix} \sim \mathcal{N}\Bigg(
\begin{bmatrix}
\bzero\\
\bzero
\end{bmatrix} \
, 
\begin{bmatrix}
\tau_{p}^{-1}I & \rho\Sigma_{OP}\\
\rho\Sigma_{OP}^{\prime} & \tau_{p}^{-1}I\\
\end{bmatrix}
\Bigg)$$
where $\rho$ is a correlation coefficient and  $\Sigma_{OP}=\tau_{o}^{-1/2}\tau_{p}^{-1/2}\bPhi_{O}^{\prime}(\bPhi_{O}\bPhi_{O}^{\prime})^{-1}L_{1}L_{2}(\bPhi_{P}\bPhi_{P}^{\prime})^{-1}\bPhi_{P}^{\prime}$, then 

$$\begin{bmatrix}
\tilde{\bW}_{O}\\
\tilde{\bW}_{P}
\end{bmatrix} \sim \mathcal{N}\Bigg(
\begin{bmatrix}
\bzero\\
\bzero
\end{bmatrix}
, 
\begin{bmatrix}
\tau_{o}^{-1/2}\bPhi_{O}\bPhi_{O}^{\prime} & \rho L_{1}L_{2}^{\prime}\\
\rho L_{2}L_{1}^{\prime} & \tau_{p}^{-1/2}\bPhi_{P}\bPhi_{P}^{\prime}\\
\end{bmatrix}
\Bigg).$$

From this, we can model the cross-covariance between the occurrence $O(\bs)$ and prevalence $P(\bs)$ spatial processes by inferring the cross-correlation between the basis coefficients $\bdel_{o}$ and $\bdel_{p}$. Under the PICAR-Z modeling framework, the dimensions of $\bdel_{o}$ and $\bdel_{p}$ are much smaller than $n$. By inferring the cross-correlation between $\bdel_{o}$ and $\bdel_{p}$, we are able to bypass the expensive matrix operations (e.g determinants and inverses) typically associated with the cross-covariance model in Equation \ref{EQ:RandomEffectsCross}. 

Using this construction, we modify the general PICAR-Z model to account for cross-correlation:

\begin{align*}
\textbf{Data Model: } & \qquad Z(\bs)|O(\bs),P(\bs) \sim F\big(\cdot|O(\bs),P(\bs)\big)\\
\textbf{Process Model: } & \qquad O(\bs)|\pi(\bs) \sim \mbox{Bern}(\pi(\bs))\\
& \qquad P(\bs)|\btheta(\bs) \sim \tilde{F}(\cdot | \btheta(\bs))\\
\textbf{Sub-process  Model 1: }& \qquad \pi(\bs)|\eta_{o}(\bs)=g^{-1}_{o}(\eta_{o}(\bs))\\
\textbf{(Occurrence) } & \qquad \eta_{o}(\bs)|\bbe_{o},\bdel_{o}(\bs)= \bX(s)^{\prime}\bbe_{o} + [\bA_{o}\bM_{o}\bdel_{o}](\bs)\\
\textbf{Sub-process  Model 2: } & \qquad \theta(\bs)|\eta_{p}(\bs)=g^{-1}_{p}(\eta_{p}(\bs))\\
\textbf{(Prevalence)} & \qquad  \eta_{p}(\bs)|\bbe_{p},\bdel_{p}(\bs)=\bX(s)^{\prime}\bbe_{p} + [\bA_{p}\bM_{p}\bdel_{p}](\bs) \\
& \qquad \begin{bmatrix}
\bdel_O\\
\bdel_P
\end{bmatrix}|\tau_o, \tau_p,\rho \sim \mathcal{N}\Bigg(
\begin{bmatrix}
\bzero\\
\bzero
\end{bmatrix} \
, 
\begin{bmatrix}
\tau_o^{-1}I & \rho\Sigma_{OP}\\
\rho\Sigma_{OP}^{\prime} & \tau_p^{-1}I\\
\end{bmatrix}
\Bigg)\\
\textbf{Parameter  Model: } & \qquad \bbe_{o}\sim \mcN(\bmu_{\beta_o}, \Sigma_{\beta_o}),\quad \bbe_{p}\sim \mcN(\bmu_{\beta_p}, \Sigma_{\beta_p})\\
& \qquad \tau_o \sim \mcG(\alpha_{\tau_o},\beta_{\tau_o}), \quad \tau_p \sim \mcG(\alpha_{\tau_p},\beta_{\tau_p})
\end{align*}
where $F\big(\cdot|O(\bs),P(\bs)\big)$ is the distribution of the specified two-part model, and $\tilde{F}(\cdot|\theta(\bs))$ denotes the distribution function of the prevalence process. $\bA_{o}$ and $\bA_{p}$ are the projectors matrices for the occurrence and prevalence processes, respectively. For the occurrence and prevalence processes, respectively, the Moran's basis functions matrices are $\mathbf{M}_{o}$ and $\mathbf{M}_{p}$, the basis coefficients are $\bdel_{o}$ and $\bdel_{p}$, and the precision parameters are $\tau_{o}$ and $\tau_{p}$. The cross-correlation coefficient is $\rho$ with cross-covariance matrix $\Sigma_{OP}=\tau_{o}^{-1/2}\tau_{p}^{-1/2}\bPhi_{O}^{\prime}(\bPhi_{O}\bPhi_{O}^{\prime})^{-1}L_{1}L_{2}(\bPhi_{P}\bPhi_{P}^{\prime})^{-1}\bPhi_{P}^{\prime}$. Similar to the PICAR-Z framework (with no cross-covariance), $\bQ$ is typically fixed prior to model fitting (see \citet{lee2021picar} for additional details). 

\section{Sensitivity Analysis on Proportion of Zeros}
We conduct a sensitivity analysis with respect to the proportion of zeros in our samples. To generate the data, we use similar settings as in the simulation study (Section 5 of the main manuscript). The proportion of zeros is set by adjusting the true values of $\bbe_o$ and $\bbe_p$, the regression coefficients of the occurrence and prevalence processes respectively. All four two-part model classes are included in the sensitivity analysis. We study three groups of dataa based on their zero composition proportions- $15-25\%$ , $40-55\%$, and $70\%+$ of the entire sample. The PICAR-Z model is fitted using the Metropolis-Hastings algorithm for 200,000 iterations. We compare metrics from the original simulation study, total rmspe, non-zero rmpse, zero vs. non-zero AUC, walltimes. In total, there are 12 different combinations in the sensitivity analysis. The sensitivity analysis results are provided in Table~\ref{Tab:Sensitivity}. We find that datasets with either a low proportion of zeros tend to have a worse classification performance than those with larger proportions of zeros. When the proportion of zeros is very high, we experience shorter model-fitting walltimes.

\begin{table}[ht]
\centering
\begin{tabular}{llrrrrr}
  \hline
  &  & Zero &  Total & non-Zero & Zero & walltime  \\
 Type & Model & Proportion & rmspe & rmspe & auc & (sec) \\ 
  \hline
count & mixture & 0.790 & 15.687 & 13.727 & 0.690 & 76.581 \\ 
  semi & mixture & 0.792 & 0.872 & 1.504 & 0.680 & 86.153 \\ 
  count & hurdle & 0.746 & 1.812 & 3.237 & 0.646 & 38.603 \\ 
  semi & hurdle & 0.742 & 1.824 & 1.840 & 0.653 & 93.806 \\ 
  count & mixture & 0.528 & 15.286 & 18.679 & 0.745 & 268.556 \\ 
  semi & mixture & 0.544 & 0.869 & 1.031 & 0.765 & 140.008 \\ 
  count & hurdle & 0.457 & 1.842 & 1.909 & 0.699 & 159.621 \\ 
  semi & hurdle & 0.441 & 1.630 & 1.719 & 0.687 & 135.610 \\ 
  count & mixture & 0.250 & 26.349 & 29.219 & 0.874 & 294.134 \\ 
  semi & mixture & 0.240 & 0.773 & 0.817 & 0.897 & 258.431 \\ 
  count & hurdle & 0.160 & 2.260 & 2.314 & 0.696 & 185.275 \\ 
  semi & hurdle & 0.169 & 2.571 & 2.754 & 0.768 & 214.443 \\ 
   \hline
\end{tabular}
 \caption{Sensitivity analysis results. The table presents the total rmspe, non-zero-valued rmspe, zero vs. non-zero AUC, walltimes (seconds), and the proportion of zero values for each dataset. }
 \label{Tab:Sensitivity}
\end{table}

\section{Applications}

\begin{table}[ht]
\centering
\begin{tabular}{|c|c|ccc|}
  \hline
  & & $\hat{\beta}_{1}$ (95\% CI) &  $\hat{\beta}_{2}$ (95\% CI) &  $\hat{\beta}_{3}$ (95\% CI) \\ 
& & Median grain size & Silt content & Altitude \\ 
  \hline
Count & PICAR-Z & -0.071 (-0.265 , 0.131) & 0.138 (-0.028 , 0.305) & 0.554 (0.459 , 0.644) \\ 
Hurdle &   PICAR-Z (Cor) & -0.093 (-0.297 , 0.102) & 0.131 (-0.033 , 0.303) & 0.557 (0.461 , 0.647) \\ 
&   Bi-square & -0.163 (-0.396 , 0.066) & -0.034 (-0.229 , 0.153) & 0.49 (0.385 , 0.595) \\ 
\hline
Count&   PICAR-Z & -0.126 (-0.361 , 0.129) & 0.009 (-0.194 , 0.207) & 0.491 (0.38 , 0.604) \\ 
Mixture &   PICAR-Z (Cor) & -0.138 (-0.374 , 0.109) & 0.008 (-0.185 , 0.207) & 0.497 (0.389 , 0.606) \\ 
&   Bi-square & -0.242 (-0.54 , 0.046) & -0.239 (-0.475 , 0.005) & 0.447 (0.32 , 0.577) \\ 
   \hline
\end{tabular}
\caption{Occurrence process parameter estimates for the bi-valve species count data example. Columns correspond to the parameter estimates and 95\% credible intervals for the regression coefficients corresponding to the three covariates (mean grain size, silt content, and altitude) for the occurrence process. Rows denote the model-fitting approach. }
\end{table}

\begin{table}[ht]
\centering
\begin{tabular}{|c|c|ccc|}
  \hline
  & & $\hat{\beta}_{1}$ (95\% CI) &  $\hat{\beta}_{2}$ (95\% CI) &  $\hat{\beta}_{3}$ (95\% CI) \\ 
& & Median grain size & Silt content & Altitude \\ 
  \hline
\multirow{2}{*}{Count} & PICAR-Z  & 0.126 (0.024 , 0.227) & 0.46 (0.38 , 0.546) & 0.229 (0.18 , 0.279) \\ 
\multirow{2}{*}{Hurdle} &   PICAR-Z (Cor) & 0.137 (0.027 , 0.235) & 0.467 (0.382 , 0.547) & 0.228 (0.178 , 0.278) \\ 
 & Bi.square & 0.11 (0.009 , 0.218) & 0.505 (0.421 , 0.589) & 0.24 (0.187 , 0.293) \\ 
  \hline
\multirow{2}{*}{Count}&   PICAR-Z & 0.11 (0.013 , 0.212) & 0.435 (0.357 , 0.513) & 0.214 (0.161 , 0.261) \\ 
\multirow{2}{*}{Mixture} &   PICAR-Z (Cor) & 0.113 (0.013 , 0.21) & 0.438 (0.353 , 0.52) & 0.212 (0.163 , 0.263) \\ 
&   Bi-square & 0.111 (0.009 , 0.206) & 0.509 (0.428 , 0.589) & 0.247 (0.196 , 0.298) \\ 
   \hline
\end{tabular}
\caption{Prevalence process parameter estimates for the bi-valve species count data example. Columns correspond to the parameter estimates and 95\% credible intervals for the regression coefficients corresponding to the three covariates (mean grain size, silt content, and altitude) for the prevalence process. Rows denote the model-fitting approach. }
\end{table}

\begin{figure}[ht]\label{Fig:Macoma}
    \centering
    \includegraphics[width=1\textwidth]{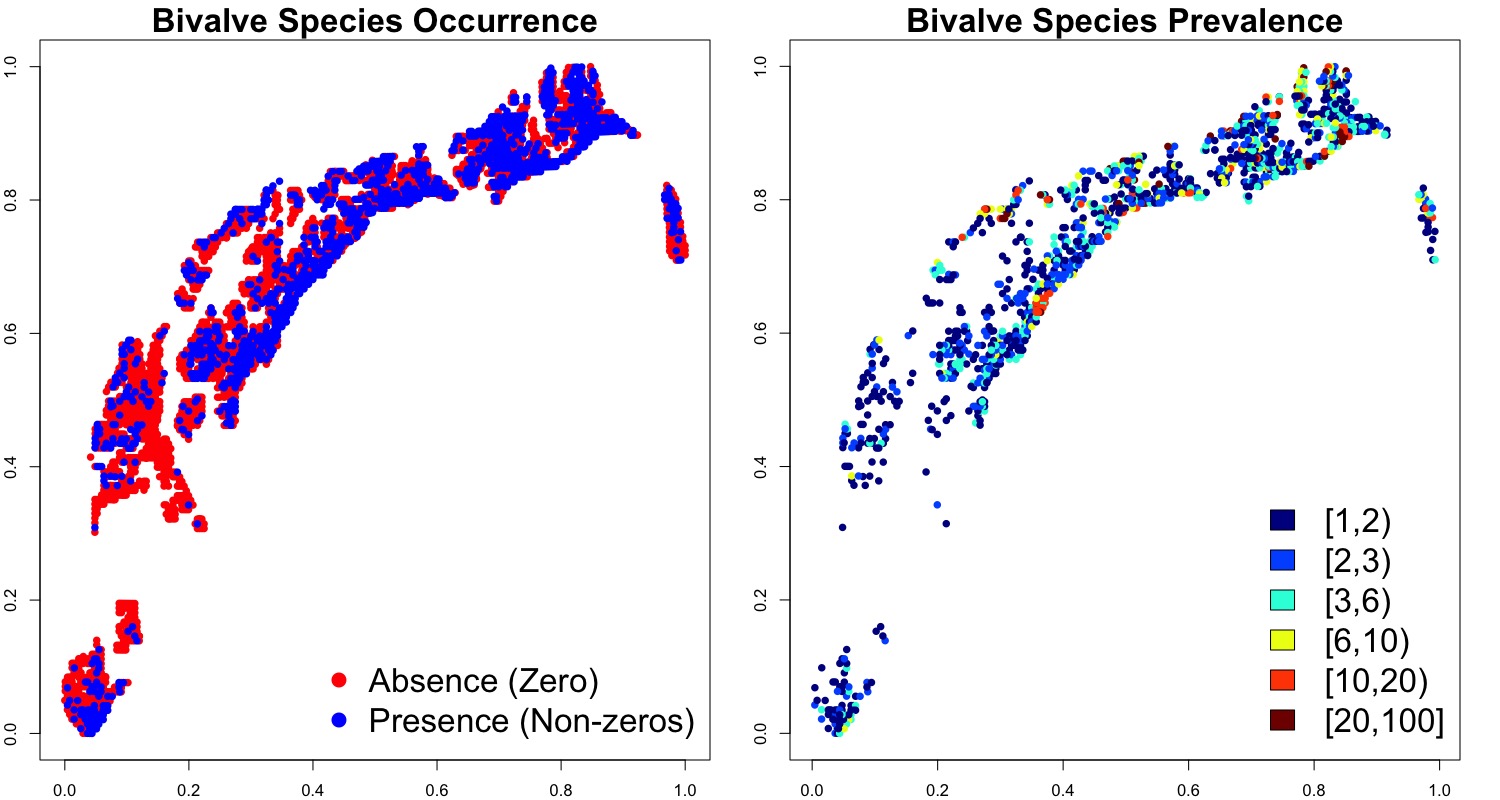}
    \caption{Maps of occurrence (left) and prevalence (right) of the Baltic tellin (Macoma balthica) species. For the occurrence map, the blue points denote the presence and the red points denote absence of the bivalve species. The prevalence map displays counts at the locations with positive counts.}
\end{figure}

\begin{figure}[h]
\begin{center}
\includegraphics[width=1\textwidth]{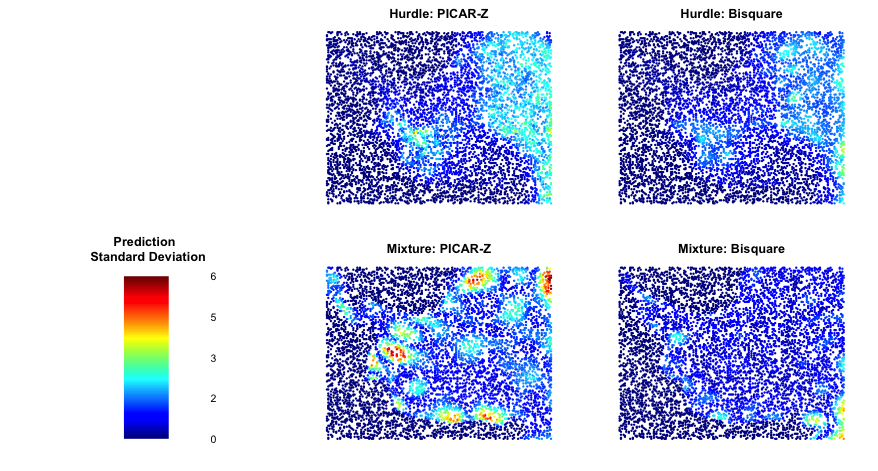} 
\caption{West Antarctic ice sheet example: Standard deviations of the predictions for the PICAR-Z and low-rank (bisquare basis) approaches.}
\label{Fig:SD_ice}
\end{center}
\end{figure}

\begin{figure}[ht]\label{Fig:SimStudySD}
    \centering
    \includegraphics[width=1\textwidth]{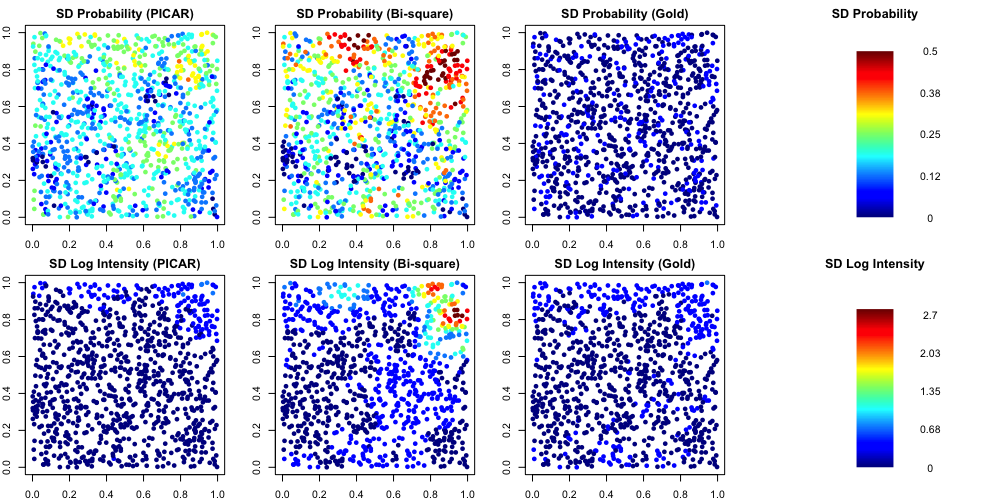}
    \caption{Prediction uncertainty for a single simulated example. Data are generated from the spatial count mixture model in the simulation study. Top row includes the standard deviation of the predicted probability surface $\pi(\bs)$ for the occurrence random process $O(\bs)$. Bottom row presents the log-intensity surface $\log(\theta(\bs))$ for the prevalence random process $P(\bs)$. Column 1 presents the true latent probability and log-intensity surfaces. The predicted surfaces for PICAR-Z (column 2), low-rank approach with bisquare basis functions (column 3), and the `reparameterized' approach (column 4) are provided. }
\end{figure}

\begin{figure}[h]
\begin{center}
\begin{tabular}{cc}
\includegraphics[width=75mm]{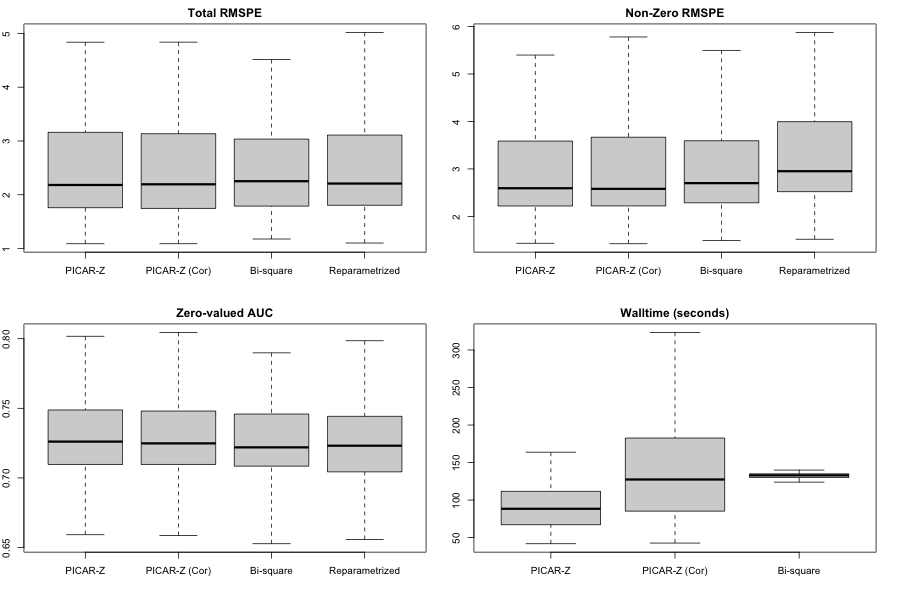} &   \includegraphics[width=75mm]{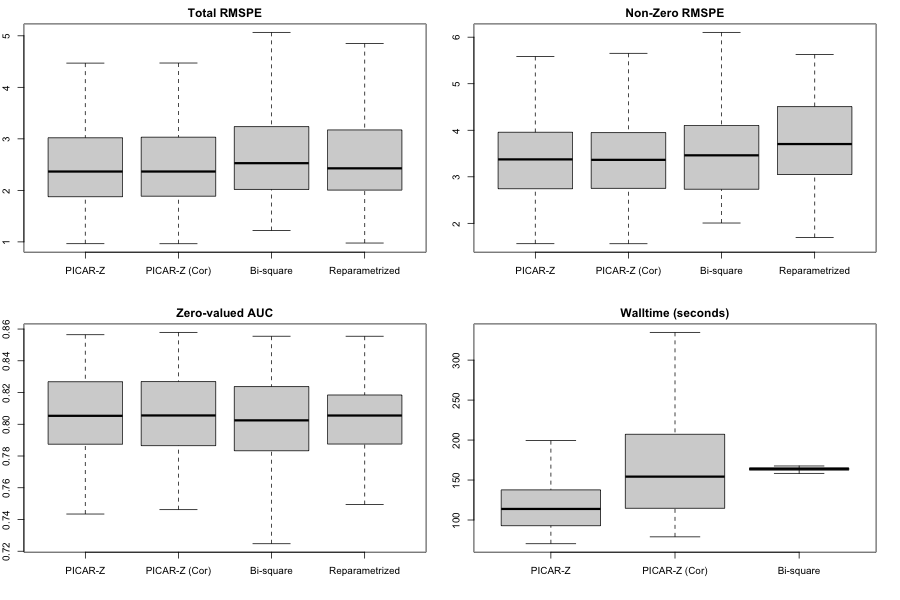} \\
(a) Count Hurdle & (b) Count Mixture\\ 
\includegraphics[width=75mm]{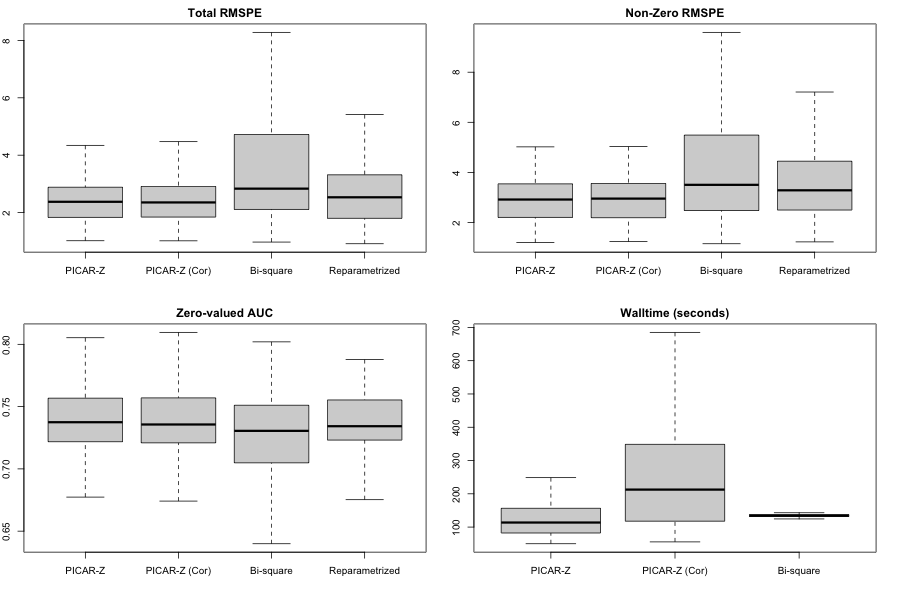} &   \includegraphics[width=75mm]{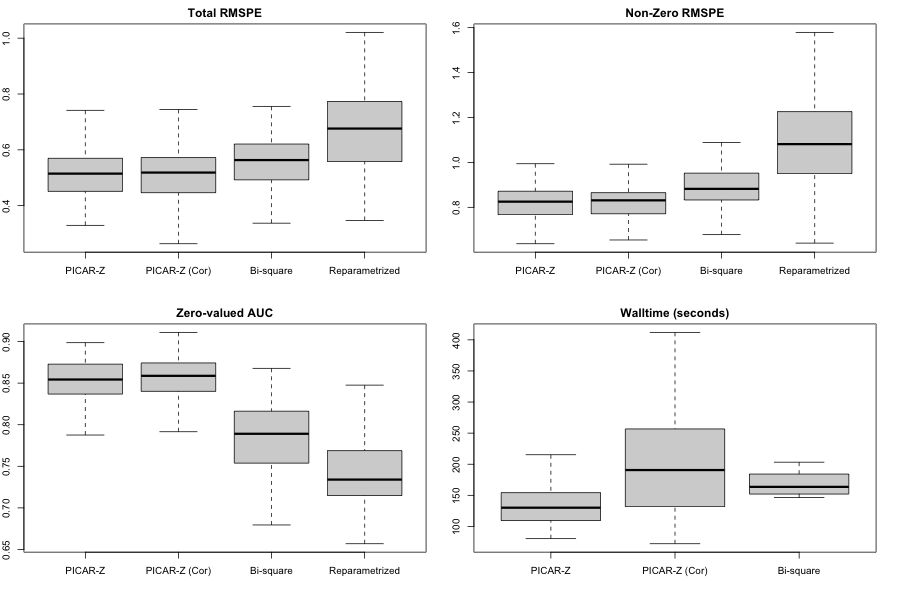} \\
(a) Semi-Continuous Hurdle & (b) Semi-Continuous Mixture\\ 
\end{tabular}
\caption{Simulation Study Results: Boxplots for Total rmspe, Non-zero rmspe, Zero-valued AUC, and walltime in seconds. (a) Count Hurdle. (b) Count Mixture. (c) Semi-Continuous Hurdle (d) Semi-Continuous Mixture.}
\label{Fig1:Partitioning}
\end{center}
\end{figure}

\begin{table}[h]
\centering
\begin{tabular}{ |c|c|c|c| } 
\hline
& & Occurrence & Prevalence\\
Class & Data Type & Median (Min , Max) & Median (Min , Max) \\
 \hline
  \multirow{2}{4em}{Hurdle} & Counts & 34.5 (3 , 116)& 49 (2 , 120)\\ 
 & Continuous & 25 (2 , 139) & 89.5 (4 , 150)\\ 
  \hline
    \multirow{2}{4em}{Mixture} & Counts & 38 (2 , 106) & 47 (2 , 109)\\ 
 & Continuous & 40 (5 , 98)& 58.5 (8 , 99)\\ 
 \hline
\end{tabular}
\caption{Rank selection results for the simulation study. Ranks are chosen via the automated heuristic. Summaries are provided for all 100 datasets in the four cases of two-part models (400 total data sets). Reported values include the median, minimum and maximum ranks for each two-part model and latent processes (occurrence and prevalence). } \label{Tab:Ranks}
\end{table}

\begin{table}[h]
\centering
\begin{tabular}{ |c|c|c| } 
\hline
 &  & Proportion Zeros \\
Class & Data Type & Mean (sd) \\
 \hline
  \multirow{2}{4em}{Hurdle} & Counts & 0.51 (0.07) \\ 
 & Continuous & 0.50 (0.07)\\ 
  \hline
    \multirow{2}{4em}{Mixture} & Counts & 0.63 (0.08) \\ 
 & Continuous & 0.66 (0.09)\\ 
 \hline
\end{tabular}
\caption{Proportion of zero-valued observations in the simulated datasets. Mean and standard deviations of the proportions for all 100 datasets in the four cases of two-part models.} 
\label{Tab:ZerosProp}
\end{table}

\newpage
\bibliography{references}

%% file: preamble.tex
\usepackage[ruled,vlined]{algorithm2e}

\newcommand\bzero{\mbox{\boldmath${0}$}}
\newcommand\bbe{\mbox{\boldmath${ \beta}$}}

\newcommand\bfeta{\mbox{\boldmath${\eta}$}}
\newcommand\bep{\mbox{\boldmath${\epsilon}$}}

\newcommand\bdel{\mbox{\boldmath${\delta}$}}
\newcommand\bmu{\mbox{\boldmath${\mu}$}}

\newcommand\btheta{\mbox{\boldmath${\theta}$}}

\newcommand\bA{{\bf A}}

\newcommand\bC{{\bf C}}

\newcommand\mcG{{\mathcal G}}

\newcommand\mcI{{\mathcal I}}

\newcommand\bL{{\bf L}}

\newcommand\bM{{\bf M}}

\newcommand\mcN{{\mathcal N}}
\newcommand\bP{{\bf P}}
\newcommand\bQ{{\bf Q}}
\newcommand\bR{{\bf R}}

\newcommand\bs{{\bf s}}

\newcommand\bv{{\bf v}}
\newcommand\bW{{\bf W}}

\newcommand\bX{{\bf X}}

\newcommand\bZ{{\bf Z}}

\makeatletter
\def\BState{\State\hskip-\ALG@thistlm}
\makeatother

%% file: manuscript.bbl
\begin{thebibliography}{}

\bibitem[Agarwal et~al., 2002]{Agarwal2002Zero}
Agarwal, D.~K., Gelfand, A.~E., and Citron-Pousty, S. (2002).
\newblock Zero-inflated models with application to spatial count data.
\newblock {\em Environmental and Ecological Statistics}, 9(4):341–355.

\bibitem[Arcuti et~al., 2016]{Arcuti2016Bayesian}
Arcuti, S., Pollice, A., Ribecco, N., and D’Onghia, G. (2016).
\newblock Bayesian spatiotemporal analysis of zero-inflated biological
  population density data by a delta-normal spatiotemporal additive model:
  Bayesian analysis of zero-inflated biological data.
\newblock {\em Biometrical Journal}, 58(2):372–386.

\bibitem[Banerjee et~al., 2013]{Banerjee2013}
Banerjee, A., Dunson, D.~B., and Tokdar, S.~T. (2013).
\newblock Efficient {G}aussian process regression for large datasets.
\newblock {\em Biometrika}, 100(1):75–89.

\bibitem[Banerjee et~al., 2008]{Banerjee2008Gaussian}
Banerjee, S., Gelfand, A.~E., Finley, A.~O., and Sang, H. (2008).
\newblock Gaussian predictive process models for large spatial data sets.
\newblock {\em Journal of the Royal Statistical Society: Series B (Statistical
  Methodology)}, 70(4):825–848.

\bibitem[Besag and Kooperberg, 1995]{besag1995conditional}
Besag, J. and Kooperberg, C. (1995).
\newblock On conditional and intrinsic autoregressions.
\newblock {\em Biometrika}, 82(4):733--746.

\bibitem[Beukema and Dekker, 2020]{beukema2020half}
Beukema, J. and Dekker, R. (2020).
\newblock Half a century of monitoring macrobenthic animals on tidal flats in
  the {D}utch {W}adden {S}ea.
\newblock {\em Marine Ecology Progress Series}, 656:1--18.

\bibitem[Bijleveld et~al., 2012]{bijleveld2012designing}
Bijleveld, A.~I., van Gils, J.~A., van~der Meer, J., Dekinga, A., Kraan, C.,
  van~der Veer, H.~W., and Piersma, T. (2012).
\newblock Designing a benthic monitoring programme with multiple conflicting
  objectives.
\newblock {\em Methods in Ecology and Evolution}, 3(3):526--536.

\bibitem[Boere and Piersma, 2012]{boere2012flyway}
Boere, G.~C. and Piersma, T. (2012).
\newblock Flyway protection and the predicament of our migrant birds: A
  critical look at international conservation policies and the dutch wadden
  sea.
\newblock {\em Ocean \& coastal management}, 68:157--168.

\bibitem[Bopp et~al., 2020]{bopp2020projecting}
Bopp, G.~P., Shaby, B.~A., Forest, C.~E., and Mej{\'\i}a, A. (2020).
\newblock Projecting flood-inducing precipitation with a {B}ayesian analogue
  model.
\newblock {\em Journal of Agricultural, Biological and Environmental
  Statistics}, 25(2):229--249.

\bibitem[Bradley et~al., 2019]{bradley2019bayesian}
Bradley, J.~R., Holan, S.~H., and Wikle, C.~K. (2019).
\newblock Bayesian hierarchical models with conjugate full-conditional
  distributions for dependent data from the natural exponential family.
\newblock {\em Journal of the American Statistical Association}, 0(ja):1--29.

\bibitem[Bradley et~al., 2015]{bradley2015spatio}
Bradley, J.~R., Wikle, C.~K., and Holan, S.~H. (2015).
\newblock Spatio-temporal change of support with application to {A}merican
  {C}ommunity {S}urvey multi-year period estimates.
\newblock {\em Stat}, 4(1):255--270.

\bibitem[Brenner and Scott, 2007]{brenner2007mathematical}
Brenner, S. and Scott, R. (2007).
\newblock {\em The {M}athematical {T}heory of {F}inite {E}lement {M}ethods},
  volume~15.
\newblock Springer Science \& Business Media.

\bibitem[Chang et~al., 2016]{chang2016calibrating}
Chang, W., Haran, M., Applegate, P., and Pollard, D. (2016).
\newblock Calibrating an ice sheet model using high-dimensional binary spatial
  data.
\newblock {\em Journal of the American Statistical Association},
  111(513):57--72.

\bibitem[Christensen and Waagepetersen, 2002]{christensen2002bayesian}
Christensen, O.~F. and Waagepetersen, R. (2002).
\newblock Bayesian prediction of spatial count data using generalized linear
  mixed models.
\newblock {\em Biometrics}, 58(2):280--286.

\bibitem[Compton et~al., 2013]{compton2013distinctly}
Compton, T.~J., Holthuijsen, S., Koolhaas, A., Dekinga, A., ten Horn, J.,
  Smith, J., Galama, Y., Brugge, M., van~der Wal, D., van~der Meer, J., et~al.
  (2013).
\newblock Distinctly variable mudscapes: {D}istribution gradients of intertidal
  macrofauna across the dutch wadden sea.
\newblock {\em Journal of Sea Research}, 82:103--116.

\bibitem[Cressie and Johannesson, 2008]{cressie2008fixed}
Cressie, N. and Johannesson, G. (2008).
\newblock Fixed rank kriging for very large spatial data sets.
\newblock {\em Journal of the Royal Statistical Society: Series B (Statistical
  Methodology)}, 70(1):209--226.

\bibitem[Cressie and Wikle, 2015]{cressie2015statistics}
Cressie, N. and Wikle, C.~K. (2015).
\newblock {\em Statistics for spatio-temporal data}.
\newblock John Wiley \& Sons.

\bibitem[Dairain et~al., 2020]{dairain2020high}
Dairain, A., Engelsma, M.~Y., Drent, J., Dekker, R., and Thieltges, D.~W.
  (2020).
\newblock High prevalences of disseminated neoplasia in the {B}altic tellin
  {L}imecola balthica in the {W}adden {S}ea.
\newblock {\em Diseases of Aquatic Organisms}, 138:89--96.

\bibitem[{de Valpine} et~al., 2017]{nimble2017}
{de Valpine}, P., Turek, D., Paciorek, C., Anderson-Bergman, C., {Temple Lang},
  D., and Bodik, R. (2017).
\newblock Programming with models: writing statistical algorithms for general
  model structures with nimble.
\newblock {\em Journal of Computational and Graphical Statistics}, 26:403--413.

\bibitem[Deschamps et~al., 2012]{deschamps2012ice}
Deschamps, P., Durand, N., Bard, E., Hamelin, B., Camoin, G., Thomas, A.~L.,
  Henderson, G.~M., Okuno, J., and Yokoyama, Y. (2012).
\newblock Ice-sheet collapse and sea-level rise at the {B{\o}lling} warming
  14,600 years ago.
\newblock {\em Nature}, 483(7391):559.

\bibitem[Diggle et~al., 1998]{diggle1998model}
Diggle, P.~J., Tawn, J.~A., and Moyeed, R. (1998).
\newblock Model-based geostatistics.
\newblock {\em Journal of the Royal Statistical Society: Series C (Applied
  Statistics)}, 47(3):299--350.

\bibitem[Dreassi et~al., 2014]{Dreassi2014Small}
Dreassi, E., Petrucci, A., and Rocco, E. (2014).
\newblock Small area estimation for semicontinuous skewed spatial data: {A}n
  application to the grape wine production in tuscany.
\newblock {\em Biometrical Journal}, 56(1):141–156.

\bibitem[Feng, 2021]{feng2021comparison}
Feng, C.~X. (2021).
\newblock A comparison of zero-inflated and hurdle models for modeling
  zero-inflated count data.
\newblock {\em Journal of Statistical Distributions and Applications},
  8(1):1--19.

\bibitem[Fernandes et~al., 2009]{Fernandes2009ST}
Fernandes, M.~V., Schmidt, A.~M., and Migon, H.~S. (2009).
\newblock Modelling zero-inflated spatio-temporal processes.
\newblock {\em Statistical Modelling: An International Journal}, 9(1):3–25.

\bibitem[Fretwell et~al., 2012]{fretwell2012bedmap2}
Fretwell, P., Pritchard, H., Vaughan, D., Bamber, J., Barrand, N., Bell, R.,
  Bianchi, C., Bingham, R., Blankenship, D., Casassa, G., et~al. (2012).
\newblock Bedmap2: {Improved} ice bed, surface and thickness datasets for
  {Antarctica}.
\newblock {\em The Cryosphere Discussions}, 6:4305--4361.

\bibitem[Fruhwirth-Schnatter and Pyne, 2010]{Fruhwirth2010}
Fruhwirth-Schnatter, S. and Pyne, S. (2010).
\newblock Bayesian inference for finite mixtures of univariate and multivariate
  skew-normal and skew-t distributions.
\newblock {\em Biostatistics}, 11(2):317–336.

\bibitem[Greve et~al., 2011]{greve2011initial}
Greve, R., Saito, F., and Abe-Ouchi, A. (2011).
\newblock Initial results of the {SeaRISE} numerical experiments with the
  models {SICOPOLIS} and {IcIES} for the greenland ice sheet.
\newblock {\em Annals of Glaciology}, 52(58):23--30.

\bibitem[Griffith, 2003]{griffith2003spatial}
Griffith, D.~A. (2003).
\newblock Spatial filtering.
\newblock In {\em Spatial Autocorrelation and Spatial Filtering}, pages
  91--130. Springer.

\bibitem[Gschl{\"o}{\ss }l and Czado, 2008]{Czado2008}
Gschl{\"o}{\ss }l, S. and Czado, C. (2008).
\newblock Modelling count data with overdispersion and spatial effects.
\newblock {\em Statistical Papers}, 49(3):531–552.

\bibitem[Guan and Haran, 2018]{Guan_Haran_2018}
Guan, Y. and Haran, M. (2018).
\newblock A computationally efficient projection-based approach for spatial
  generalized linear mixed models.
\newblock {\em Journal of Computational and Graphical Statistics},
  27(4):701–714.

\bibitem[Guan and Haran, 2019]{guan2019fast}
Guan, Y. and Haran, M. (2019).
\newblock Fast expectation-maximization algorithms for spatial generalized
  linear mixed models.
\newblock {\em arXiv preprint arXiv:1909.05440}.

\bibitem[Guhaniyogi et~al., 2011]{guhaniyogi2011adaptive}
Guhaniyogi, R., Finley, A.~O., Banerjee, S., and Gelfand, A.~E. (2011).
\newblock Adaptive {G}aussian predictive process models for large spatial
  datasets.
\newblock {\em Environmetrics}, 22(8):997--1007.

\bibitem[Haran, 2011]{haran2011gaussian}
Haran, M. (2011).
\newblock Gaussian random field models for spatial data.
\newblock {\em Handbook of Markov Chain Monte Carlo}, pages 449--478.

\bibitem[Haran et~al., 2003]{haran2003accelerating}
Haran, M., Hodges, J.~S., and Carlin, B.~P. (2003).
\newblock Accelerating computation in markov random field models for spatial
  data via structured {MCMC}.
\newblock {\em Journal of Computational and Graphical Statistics},
  12(2):249--264.

\bibitem[Higdon, 1998]{higdon1998process}
Higdon, D. (1998).
\newblock A process-convolution approach to modelling temperatures in the
  {N}orth {A}tlantic {O}cean.
\newblock {\em Environmental and Ecological Statistics}, 5(2):173--190.

\bibitem[Hjelle and D{\ae}hlen, 2006]{hjelle2006triangulations}
Hjelle, {\O}. and D{\ae}hlen, M. (2006).
\newblock {\em Triangulations and applications}.
\newblock Springer Science \& Business Media.

\bibitem[Hughes and Haran, 2013]{hughes2013dimension}
Hughes, J. and Haran, M. (2013).
\newblock Dimension reduction and alleviation of confounding for spatial
  generalized linear mixed models.
\newblock {\em Journal of the Royal Statistical Society: Series B (Statistical
  Methodology)}, 75(1):139--159.

\bibitem[Katzfuss, 2017]{katzfuss2017multi}
Katzfuss, M. (2017).
\newblock A multi-resolution approximation for massive spatial datasets.
\newblock {\em Journal of the American Statistical Association},
  112(517):201--214.

\bibitem[Khan and Calder, 2022]{khan2022restricted}
Khan, K. and Calder, C.~A. (2022).
\newblock Restricted spatial regression methods: Implications for inference.
\newblock {\em Journal of the American Statistical Association},
  117(537):482--494.

\bibitem[Kim et~al., 2012]{kim2012blup}
Kim, S.~H., Chang, C.-C.~H., Kim, K.~H., Fine, M.~J., and Stone, R.~A. (2012).
\newblock {BLUP (REMQL)} estimation of a correlated random effects negative
  binomial hurdle model.
\newblock {\em Health Services and Outcomes Research Methodology},
  12(4):302--319.

\bibitem[Kleiber and Nychka, 2012]{kleiber2012nonstationary}
Kleiber, W. and Nychka, D. (2012).
\newblock Nonstationary modeling for multivariate spatial processes.
\newblock {\em Journal of Multivariate Analysis}, 112:76--91.

\bibitem[Krock et~al., 2021]{krock2021modeling}
Krock, M., Kleiber, W., Hammerling, D., and Becker, S. (2021).
\newblock Modeling massive highly-multivariate nonstationary spatial data with
  the basis graphical lasso.
\newblock {\em arXiv preprint arXiv:2101.02404}.

\bibitem[Lambert, 1992]{lambert1992zero}
Lambert, D. (1992).
\newblock Zero-inflated {Poisson} regression, with an application to defects in
  manufacturing.
\newblock {\em Technometrics}, 34(1):1--14.

\bibitem[Lee and Haran, 2022]{lee2021picar}
Lee, B.~S. and Haran, M. (2022).
\newblock {PICAR:} an efficient extendable approach for fitting hierarchical
  spatial models.
\newblock {\em Technometrics}, 64(2):187--198.

\bibitem[Lee and Kim, 2017]{lee2017applicability}
Lee, C.-E. and Kim, S. (2017).
\newblock Applicability of zero-inflated models to fit the torrential rainfall
  count data with extra zeros in south korea.
\newblock {\em Water}, 9(2):123.

\bibitem[Lee et~al., 2016]{lee2016spatial}
Lee, Y., Alam, M.~M., Noh, M., R{\"o}nneg{\aa}rd, L., and Skarin, A. (2016).
\newblock Spatial modeling of data with excessive zeros applied to reindeer
  pellet-group counts.
\newblock {\em Ecology and evolution}, 6(19):7047--7056.

\bibitem[Lehoucq et~al., 1998]{lehoucq1998arpack}
Lehoucq, R.~B., Sorensen, D.~C., and Yang, C. (1998).
\newblock {\em {ARPACK} users' guide: {S}olution of large-scale eigenvalue
  problems with implicitly restarted Arnoldi methods}, volume~6.
\newblock Siam.

\bibitem[Lindgren et~al., 2015]{lindgren2015bayesian}
Lindgren, F., Rue, H., et~al. (2015).
\newblock Bayesian spatial modelling with {R-INLA}.
\newblock {\em Journal of Statistical Software}, 63(19):1--25.

\bibitem[Lindgren et~al., 2011]{Lindgren2011}
Lindgren, F., Rue, H., and Lindström, J. (2011).
\newblock An explicit link between {G}aussian fields and {G}aussian {M}arkov
  random fields: the stochastic partial differential equation approach.
\newblock {\em Journal of the {R}oyal {S}tatistical {S}ociety: {S}eries {B}
  (Statistical Methodology)}, 73(4):423--498.

\bibitem[Liu et~al., 2016]{liu2016analyzing}
Liu, L., Strawderman, R.~L., Johnson, B.~A., and O'Quigley, J.~M. (2016).
\newblock Analyzing repeated measures semi-continuous data, with application to
  an alcohol dependence study.
\newblock {\em Statistical Methods in Medical Research}, 25(1):133--152.

\bibitem[Lyashevska et~al., 2016]{lyashevska2016mapping}
Lyashevska, O., Brus, D.~J., and van~der Meer, J. (2016).
\newblock Mapping species abundance by a spatial zero-inflated {P}oisson model:
  {A} case study in the {W}adden {S}ea, the {N}etherlands.
\newblock {\em Ecology and {E}volution}, 6(2):532--543.

\bibitem[McGranahan et~al., 2007]{mcgranahan2007rising}
McGranahan, G., Balk, D., and Anderson, B. (2007).
\newblock The rising tide: {A}ssessing the risks of climate change and human
  settlements in low elevation coastal zones.
\newblock {\em Environment and Urbanization}, 19(1):17--37.

\bibitem[Min and Agresti, 2005]{min2005random}
Min, Y. and Agresti, A. (2005).
\newblock Random effect models for repeated measures of zero-inflated count
  data.
\newblock {\em Statistical modelling}, 5(1):1--19.

\bibitem[Moran, 1950]{moran1950notes}
Moran, P.~A. (1950).
\newblock Notes on continuous stochastic phenomena.
\newblock {\em Biometrika}, 37(1/2):17--23.

\bibitem[Mullahy, 1986]{Mullahy1986}
Mullahy, J. (1986).
\newblock Specification and testing of some modified count data models.
\newblock {\em Journal of Econometrics}, 33(3):341–365.

\bibitem[Neelon, 2019]{neelon2019bayesian}
Neelon, B. (2019).
\newblock Bayesian zero-inflated negative binomial regression based on
  {P}{\'o}lya-gamma mixtures.
\newblock {\em Bayesian {A}nalysis}, 14(3):829.

\bibitem[Neelon et~al., 2016a]{Neelon2016Spatiotemporal}
Neelon, B., Chang, H.~H., Ling, Q., and Hastings, N.~S. (2016a).
\newblock Spatiotemporal hurdle models for zero-inflated count data: Exploring
  trends in emergency department visits.
\newblock {\em Statistical Methods in Medical Research}, 25(6):2558–2576.

\bibitem[Neelon et~al., 2016b]{neelon2016modeling}
Neelon, B., O'Malley, A.~J., and Smith, V.~A. (2016b).
\newblock Modeling zero-modified count and semicontinuous data in health
  services research {P}art 1: {B}ackground and overview.
\newblock {\em Statistics in Medicine}, 35(27):5070--5093.

\bibitem[Neelon et~al., 2011]{Neelon2011Expenditure}
Neelon, B., O’Malley, A.~J., and Normand, S.-L.~T. (2011).
\newblock A {B}ayesian two-part latent class model for longitudinal medical
  expenditure data: Assessing the impact of mental health and substance abuse
  parity.
\newblock {\em Biometrics}, 67(1):280–289.

\bibitem[Neelon et~al., 2015]{Neelon2015Semicontinuous}
Neelon, B., Zhu, L., and Neelon, S. E.~B. (2015).
\newblock Bayesian two-part spatial models for semicontinuous data with
  application to emergency department expenditures.
\newblock {\em Biostatistics}, 16(3):465–479.

\bibitem[Nicholls et~al., 2008]{nicholls2008global}
Nicholls, R.~J., Tol, R.~S., and Vafeidis, A.~T. (2008).
\newblock Global estimates of the impact of a collapse of the {West Antarctic}
  ice sheet: {An} application of {FUND}.
\newblock {\em Climatic Change}, 91(1-2):171.

\bibitem[Nychka et~al., 2015]{nychka2015multiresolution}
Nychka, D., Bandyopadhyay, S., Hammerling, D., Lindgren, F., and Sain, S.
  (2015).
\newblock A multiresolution {G}aussian process model for the analysis of large
  spatial datasets.
\newblock {\em Journal of Computational and Graphical Statistics},
  24(2):579--599.

\bibitem[Oliver, 2003]{oliver2003gaussian}
Oliver, D.~S. (2003).
\newblock Gaussian cosimulation: modelling of the cross-covariance.
\newblock {\em Mathematical Geology}, 35(6):681--698.

\bibitem[Olsen and Schafer, 2001]{Olsen2001Two}
Olsen, M.~K. and Schafer, J.~L. (2001).
\newblock A two-part random-effects model for semicontinuous longitudinal data.
\newblock {\em Journal of the American Statistical Association},
  96(454):730–745.

\bibitem[Park and Haran, 2020]{park2020reduced}
Park, J. and Haran, M. (2020).
\newblock Reduced-dimensional {M}onte {C}arlo maximum likelihood for latent
  {G}aussian random field models.
\newblock {\em Journal of Computational and Graphical Statistics},
  30(2):269--283.

\bibitem[Qiu and Mei, 2019]{RSpectra2019}
Qiu, Y. and Mei, J. (2019).
\newblock {\em RSpectra: {S}olvers for Large-Scale Eigenvalue and {SVD}
  Problems}.
\newblock R package version 0.15-0.

\bibitem[Rasmussen and Williams, 2006]{williams2006gaussian}
Rasmussen, C. and Williams, C. (2006).
\newblock {\em {G}aussian {P}rocesses for {M}achine {L}earning}.
\newblock Adaptive Computation and Machine Learning. MIT Press, Cambridge, MA,
  USA.

\bibitem[Rasmussen, 2004]{rasmussen2004gaussian}
Rasmussen, C.~E. (2004).
\newblock Gaussian processes in machine learning.
\newblock In {\em Advanced {L}ectures on {M}achine {L}earning}, pages 63--71.
  Springer.

\bibitem[Rathbun and Fei, 2006]{Rathbun2006Spatial}
Rathbun, S.~L. and Fei, S. (2006).
\newblock A spatial zero-inflated poisson regression model for oak
  regeneration.
\newblock {\em Environmental and Ecological Statistics}, 13(4):409–426.

\bibitem[Recta et~al., 2012]{recta2012two}
Recta, V., Haran, M., and Rosenberger, J.~L. (2012).
\newblock A two-stage model for incidence and prevalence in point-level spatial
  count data.
\newblock {\em Environmetrics}, 23(2):162--174.

\bibitem[Roeder et~al., 1999]{roeder1999modeling}
Roeder, K., Lynch, K.~G., and Nagin, D.~S. (1999).
\newblock Modeling uncertainty in latent class membership: {A} case study in
  criminology.
\newblock {\em Journal of the American Statistical Association},
  94(447):766--776.

\bibitem[Rue et~al., 2009]{rue2009approximate}
Rue, H., Martino, S., and Chopin, N. (2009).
\newblock Approximate {B}ayesian inference for latent {G}aussian models by
  using integrated nested {L}aplace approximations.
\newblock {\em Journal of the {R}oyal {S}tatistical {S}ociety: Series {B}
  (Statistical Methodology)}, 71(2):319--392.

\bibitem[Sengupta and Cressie, 2013]{sengupta2013hierarchical}
Sengupta, A. and Cressie, N. (2013).
\newblock Hierarchical statistical modeling of big spatial datasets using the
  exponential family of distributions.
\newblock {\em Spatial Statistics}, 4:14--44.

\bibitem[Serreze and Barry, 2011]{serreze2011processes}
Serreze, M.~C. and Barry, R.~G. (2011).
\newblock Processes and impacts of {A}rctic amplification: {A} research
  synthesis.
\newblock {\em Global and planetary change}, 77(1-2):85--96.

\bibitem[Stein, 2012]{stein2012interpolation}
Stein, M.~L. (2012).
\newblock {\em Interpolation of Spatial Data: Some Theory for Kriging}.
\newblock Springer Science \& Business Media.

\bibitem[van~der Meer et~al., 2003]{van2003large}
van~der Meer, J., Beukema, J.~J., and Dekker, R. (2003).
\newblock Large spatial variability in lifetime egg production in an intertidal
  {B}altic tellin ({M}acoma balthica) population.
\newblock {\em Helgoland Marine Research}, 56:274--278.

\bibitem[Ver~Hoef and Jansen, 2007]{Hoef2007Seals}
Ver~Hoef, J.~M. and Jansen, J.~K. (2007).
\newblock Space—time zero‐inflated count models of harbor seals.
\newblock {\em Environmetrics}, 18(7):697–712.

\bibitem[Vuong, 1989]{vuong1989likelihood}
Vuong, Q.~H. (1989).
\newblock Likelihood ratio tests for model selection and non-nested hypotheses.
\newblock {\em Econometrica: Journal of the Econometric Society}, pages
  307--333.

\bibitem[Wang et~al., 2015]{Wang2014}
Wang, X., Chen, M.-H., Kuo, R.~C., and Dey, D.~K. (2015).
\newblock Bayesian spatial-temporal modeling of ecological zero-inflated count
  data.
\newblock {\em Statistica Sinica}, 25(1):189.

\bibitem[Zhang, 2007]{zhang2007increasing}
Zhang, J. (2007).
\newblock Increasing {A}ntarctic sea ice under warming atmospheric and oceanic
  conditions.
\newblock {\em Journal of Climate}, 20(11):2515--2529.

\bibitem[Zimmerman and Ver~Hoef, 2022]{zimmerman2022deconfounding}
Zimmerman, D.~L. and Ver~Hoef, J.~M. (2022).
\newblock On deconfounding spatial confounding in linear models.
\newblock {\em The American Statistician}, 76(2):159--167.

\end{thebibliography}


\begin{thebibliography}{}

\bibitem[Agarwal et~al., 2002]{Agarwal2002Zero}
Agarwal, D.~K., Gelfand, A.~E., and Citron-Pousty, S. (2002).
\newblock Zero-inflated models with application to spatial count data.
\newblock {\em Environmental and Ecological Statistics}, 9(4):341–355.

\bibitem[Chai and Bailey, 2008]{chai2008use}
Chai, H.~S. and Bailey, K.~R. (2008).
\newblock Use of log-skew-normal distribution in analysis of continuous data
  with a discrete component at zero.
\newblock {\em Statistics in medicine}, 27(18):3643--3655.

\bibitem[Christensen et~al., 2006]{Christensen2006robust}
Christensen, O.~F., Roberts, G.~O., and Sk{\"o}ld, M. (2006).
\newblock Robust {M}arkov chain {M}onte {C}arlo methods for spatial generalized
  linear mixed models.
\newblock {\em Journal of Computational and Graphical Statistics}, 15(1):1--17.

\bibitem[Christensen and Waagepetersen, 2002]{christensen2002bayesian}
Christensen, O.~F. and Waagepetersen, R. (2002).
\newblock Bayesian prediction of spatial count data using generalized linear
  mixed models.
\newblock {\em Biometrics}, 58(2):280--286.

\bibitem[Cressie and Johannesson, 2008]{cressie2008fixed}
Cressie, N. and Johannesson, G. (2008).
\newblock Fixed rank kriging for very large spatial data sets.
\newblock {\em Journal of the Royal Statistical Society: Series B (Statistical
  Methodology)}, 70(1):209--226.

\bibitem[Cressie and Kang, 2010]{cressie2010high}
Cressie, N. and Kang, E.~L. (2010).
\newblock High-resolution digital soil mapping: Kriging for very large
  datasets.
\newblock In {\em Proximal Soil Sensing}, pages 49--63. Springer.

\bibitem[Diggle et~al., 1998]{diggle1998model}
Diggle, P.~J., Tawn, J.~A., and Moyeed, R. (1998).
\newblock Model-based geostatistics.
\newblock {\em Journal of the Royal Statistical Society: Series C (Applied
  Statistics)}, 47(3):299--350.

\bibitem[Dreassi et~al., 2014]{Dreassi2014Small}
Dreassi, E., Petrucci, A., and Rocco, E. (2014).
\newblock Small area estimation for semicontinuous skewed spatial data: {A}n
  application to the grape wine production in tuscany.
\newblock {\em Biometrical Journal}, 56(1):141–156.

\bibitem[Fruhwirth-Schnatter and Pyne, 2010]{Fruhwirth2010}
Fruhwirth-Schnatter, S. and Pyne, S. (2010).
\newblock Bayesian inference for finite mixtures of univariate and multivariate
  skew-normal and skew-t distributions.
\newblock {\em Biostatistics}, 11(2):317–336.

\bibitem[Gschl{\"o}{\ss }l and Czado, 2008]{Czado2008}
Gschl{\"o}{\ss }l, S. and Czado, C. (2008).
\newblock Modelling count data with overdispersion and spatial effects.
\newblock {\em Statistical Papers}, 49(3):531–552.

\bibitem[Guan and Haran, 2018]{Guan_Haran_2018}
Guan, Y. and Haran, M. (2018).
\newblock A computationally efficient projection-based approach for spatial
  generalized linear mixed models.
\newblock {\em Journal of Computational and Graphical Statistics},
  27(4):701–714.

\bibitem[Hall, 2000]{hall2000zero}
Hall, D.~B. (2000).
\newblock Zero-inflated poisson and binomial regression with random effects:
  {A} case study.
\newblock {\em Biometrics}, 56(4):1030--1039.

\bibitem[Haran, 2011]{haran2011gaussian}
Haran, M. (2011).
\newblock Gaussian random field models for spatial data.
\newblock {\em Handbook of Markov Chain Monte Carlo}, pages 449--478.

\bibitem[Haran et~al., 2003]{haran2003accelerating}
Haran, M., Hodges, J.~S., and Carlin, B.~P. (2003).
\newblock Accelerating computation in markov random field models for spatial
  data via structured {MCMC}.
\newblock {\em Journal of Computational and Graphical Statistics},
  12(2):249--264.

\bibitem[Lambert, 1992]{lambert1992zero}
Lambert, D. (1992).
\newblock Zero-inflated {Poisson} regression, with an application to defects in
  manufacturing.
\newblock {\em Technometrics}, 34(1):1--14.

\bibitem[Lee and Haran, 2022]{lee2021picar}
Lee, B.~S. and Haran, M. (2022).
\newblock {PICAR:} an efficient extendable approach for fitting hierarchical
  spatial models.
\newblock {\em Technometrics}, 64(2):187--198.

\bibitem[Liu et~al., 2016]{liu2016analyzing}
Liu, L., Strawderman, R.~L., Johnson, B.~A., and O'Quigley, J.~M. (2016).
\newblock Analyzing repeated measures semi-continuous data, with application to
  an alcohol dependence study.
\newblock {\em Statistical Methods in Medical Research}, 25(1):133--152.

\bibitem[Moulton and Halsey, 1995]{moulton1995mixture}
Moulton, L.~H. and Halsey, N.~A. (1995).
\newblock A mixture model with detection limits for regression analyses of
  antibody response to vaccine.
\newblock {\em Biometrics}, pages 1570--1578.

\bibitem[Mullahy, 1986]{Mullahy1986}
Mullahy, J. (1986).
\newblock Specification and testing of some modified count data models.
\newblock {\em Journal of Econometrics}, 33(3):341–365.

\bibitem[Mwalili et~al., 2008]{mwalili2008zero}
Mwalili, S.~M., Lesaffre, E., and Declerck, D. (2008).
\newblock The zero-inflated negative binomial regression model with correction
  for misclassification: an example in caries research.
\newblock {\em Statistical methods in medical research}, 17(2):123--139.

\bibitem[Neelon et~al., 2013]{Neelon2013Poisson}
Neelon, B., Ghosh, P., and Loebs, P.~F. (2013).
\newblock A spatial poisson hurdle model for exploring geographic variation in
  emergency department visits: Spatial hurdle model for exploring geographic
  variation.
\newblock {\em Journal of the Royal Statistical Society: Series A (Statistics
  in Society)}, 176(2):389–413.

\bibitem[Neelon et~al., 2016]{neelon2016modeling}
Neelon, B., O'Malley, A.~J., and Smith, V.~A. (2016).
\newblock Modeling zero-modified count and semicontinuous data in health
  services research {P}art 1: {B}ackground and overview.
\newblock {\em Statistics in Medicine}, 35(27):5070--5093.

\bibitem[Neelon et~al., 2015]{Neelon2015Semicontinuous}
Neelon, B., Zhu, L., and Neelon, S. E.~B. (2015).
\newblock Bayesian two-part spatial models for semicontinuous data with
  application to emergency department expenditures.
\newblock {\em Biostatistics}, 16(3):465–479.

\bibitem[Oliver, 2003]{oliver2003gaussian}
Oliver, D.~S. (2003).
\newblock Gaussian cosimulation: modelling of the cross-covariance.
\newblock {\em Mathematical Geology}, 35(6):681--698.

\bibitem[Rathbun and Fei, 2006]{Rathbun2006Spatial}
Rathbun, S.~L. and Fei, S. (2006).
\newblock A spatial zero-inflated poisson regression model for oak
  regeneration.
\newblock {\em Environmental and Ecological Statistics}, 13(4):409–426.

\bibitem[Recta et~al., 2012]{recta2012two}
Recta, V., Haran, M., and Rosenberger, J.~L. (2012).
\newblock A two-stage model for incidence and prevalence in point-level spatial
  count data.
\newblock {\em Environmetrics}, 23(2):162--174.

\bibitem[Sengupta and Cressie, 2013]{sengupta2013hierarchical}
Sengupta, A. and Cressie, N. (2013).
\newblock Hierarchical statistical modeling of big spatial datasets using the
  exponential family of distributions.
\newblock {\em Spatial Statistics}, 4:14--44.

\bibitem[Sengupta et~al., 2016]{sengupta2016predictive}
Sengupta, A., Cressie, N., Kahn, B.~H., and Frey, R. (2016).
\newblock Predictive inference for big, spatial, non-{G}aussian data: {MODIS}
  cloud data and its change-of-support.
\newblock {\em Australian \& New Zealand Journal of Statistics}, 58(1):15--45.

\bibitem[Shi and Kang, 2017]{shi2017spatial}
Shi, H. and Kang, E.~L. (2017).
\newblock Spatial data fusion for large non-{G}aussian remote sensing datasets.
\newblock {\em Stat}, 6(1):390--404.

\bibitem[Stein, 2012]{stein2012interpolation}
Stein, M.~L. (2012).
\newblock {\em Interpolation of Spatial Data: Some Theory for Kriging}.
\newblock Springer Science \& Business Media.

\bibitem[Tobin, 1958]{tobin1958estimation}
Tobin, J. (1958).
\newblock Estimation of relationships for limited dependent variables.
\newblock {\em Econometrica: Journal of the Econometric Society}, pages 24--36.

\bibitem[Ver~Hoef and Jansen, 2007]{Hoef2007Seals}
Ver~Hoef, J.~M. and Jansen, J.~K. (2007).
\newblock Space—time zero‐inflated count models of harbor seals.
\newblock {\em Environmetrics}, 18(7):697–712.

\end{thebibliography}
